\documentclass[
	aps, prl, superscriptaddress, twocolumn,
	10pt
	floatfix, 
    nofootinbib,
	tightenlines
]{revtex4-1}
\usepackage[final]{graphicx}
\usepackage{times,bbm,amsmath,amssymb}
\usepackage{epsfig,color}
\usepackage{xcolor}
\usepackage{hyperref}
\hypersetup{
    colorlinks = true
}
\usepackage{cleveref}

\usepackage{float,siunitx}
\usepackage[caption = false]{subfig}

\usepackage[greek,english]{babel}
\usepackage{thumbpdf,enumerate}
\usepackage{booktabs}
\usepackage{sidecap}
\usepackage[scaled=.8]{couriers}    
\usepackage{pstricks}
\usepackage{multirow}
\usepackage{placeins}
 \usepackage{relsize}  
\usepackage{pst-grad,bm}
\usepackage{epigraph}
\usepackage{gensymb}
\usepackage{longtable}
\usepackage{soul}
\usepackage{ulem} 
\usepackage{acronym}
\usepackage{physics}
\usepackage{easyReview}

\newacro{ML}{Machine Learning}
\newacro{VVB}{Vector Vortex Beam}
\newacro{QW}{Quantum Walk}
\newacro{CNN}{Convolutional Neural Network}
\newacro{NN}{Neural Network}
\newacro{LG}{Laguerre-Gauss}
\newacro{HG}{Hermite-Gauss}
\newacro{OAM}{Orbital Angular Momentum}
\newacro{SAM}{Spin Angular Momentum}
\newacro{CCD}{Charged-Coupled Device}
\newacro{SVM}{Support Vector Machine}
\newacro{DR}{Dimensionality Reduction}
\newacro{PCA}{Principal Component Analysis}
\newacro{RBF}{Radial Basis Function}

\begin{document}


\title{Machine learning-based classification of vector vortex beams}

\author{Taira Giordani}
\affiliation{Dipartimento di Fisica, Sapienza Universit\`{a} di Roma,
Piazzale Aldo Moro 5, I-00185 Roma, Italy}

\author{Alessia Suprano}
\affiliation{Dipartimento di Fisica, Sapienza Universit\`{a} di Roma,
Piazzale Aldo Moro 5, I-00185 Roma, Italy}

\author{Emanuele Polino}
\affiliation{Dipartimento di Fisica, Sapienza Universit\`{a} di Roma,
Piazzale Aldo Moro 5, I-00185 Roma, Italy}
\author{Francesca Acanfora}
\affiliation{Dipartimento di Fisica, Sapienza Universit\`{a} di Roma,
Piazzale Aldo Moro 5, I-00185 Roma, Italy}

\author{Luca Innocenti}
\affiliation{Centre for Theoretical Atomic, Molecular, and Optical Physics,
School of Mathematics and Physics, Queen's University Belfast, BT7 1NN Belfast, United Kingdom}

\author{Alessandro Ferraro}
\affiliation{Centre for Theoretical Atomic, Molecular, and Optical Physics,
School of Mathematics and Physics, Queen's University Belfast, BT7 1NN Belfast, United Kingdom}

\author{Mauro Paternostro}
\affiliation{Centre for Theoretical Atomic, Molecular, and Optical Physics,
School of Mathematics and Physics, Queen's University Belfast, BT7 1NN Belfast, United Kingdom}

\author{Nicol\`o Spagnolo}
\affiliation{Dipartimento di Fisica, Sapienza Universit\`{a} di Roma,
Piazzale Aldo Moro 5, I-00185 Roma, Italy}

\author{Fabio Sciarrino}
\affiliation{Dipartimento di Fisica, Sapienza Universit\`{a} di Roma,
Piazzale Aldo Moro 5, I-00185 Roma, Italy}
\affiliation{Consiglio Nazionale delle Ricerche, Istituto dei sistemi Complessi (CNR-ISC), Via dei Taurini 19, 00185 Roma, Italy}

\begin{abstract}
    Structured light is attracting significant attention for its diverse applications in both classical and quantum optics. The so-called vector vortex beams display peculiar properties in both contexts due to the non-trivial correlations between optical polarization and orbital angular momentum.
    Here we {demonstrate} a new, flexible {experimental approach} to the classification of vortex vector beams.
    We first describe a platform for generating arbitrary complex vector vortex beams inspired to photonic quantum walks. We then exploit recent machine learning methods -- namely convolutional neural networks and principal component analysis -- to recognize and classify specific polarization patterns. Our study demonstrates the significant advantages resulting from the use of machine learning-based protocols for the construction and characterization of high-dimensional resources for quantum protocols.
\end{abstract}
\maketitle

\textit{Introduction--} Light is endowed with~\ac{OAM}~\cite{allen_0AM_1992,padgett2004light}, a degree of freedom associated with structured, non-plane wavefronts, and characterized by an azimuthal phase dependence. 
When a nontrivial phase dependence is coupled with a helicoidal transverse polarization pattern, one talks of a \ac{VVB}~\cite{erhard2018twisted, padgett2004light}.
The interest in such states is motivated by the applications in multiple fields of classical and quantum optics~\cite{Marrucci2011Rev, Cozzolino_rev}: from particle trapping to metrological applications in microscopy~\cite{Cardano2015Rev, roadMap}, and for OAM-based communications schemes in free-space and in-fibre~\cite{Willner:15,cozzolino2019air}.
\acp{VVB} are also often employed in quantum information protocols due to the hyperentanglement between their polarization and spatial degrees of freedom. Photonic platforms for quantum sensing and metrology leveraging such encoding have also been reported~\cite{fickler2012quantum,dambrosio_gear2013}. OAM-based schemes for investigating quantum causal structures~\cite{Goswami2018}, quantum communication and cryptography~\cite{vallone_qkd_2014,Wang2015,Mirhosseini_2015,Malik2016,Sit17,Cozzolino2019_fiber}, quantum walks~\cite{zhang-oam-qw-2010,goyal2013implementing,cardano2015quantum}, quantum simulation~\cite{cardano2016statistical,cardano_zak_2017}, and quantum state engineering~\cite{Innocenti2017,giordani_2018}, have been previously demonstrated. 

Despite the potential of \acp{VVB}, many questions regarding the decoding of information stored in OAM and polarization remain unanswered. Various techniques of OAM-demultiplexing envisage the need of additional instruments -- such as interferometry~\cite{Leach2002_oamSorter,Slussarenko2010_oamSorter,Bauer2014} or spatial filtering~\cite{Berkhout2010_oamSorter,bolduc2013holo,Mehul2014_oamSorter} -- to be efficiently implemented.
These introduce detrimental effects of loss and noise~\cite{Qassim2014}. Moreover, the challenge of performing state tomography in such a high-dimensional framework, a fundamental task in quantum information processing~\cite{Paris2004, Banaszek_2013}, can hardly be overestimated. The design and demonstration of reliable techniques for the generation and classification of~\acp{VVB} is thus highly desirable. Indeed, substantive efforts on finding novel platforms are subject of intense research activities~\cite{Liu:17,Ndagano:18, Cardano2015Rev,roadMap},
including in integrated photonics~\cite{ oamchip,cai2012integrated,chiptofiber} and generation by plasmonic metasurfaces~\cite{karimi2014,yue2016}.

Recently,~\ac{ML} has emerged as a versatile toolbox to tackle a variety of tasks arising in experimental platforms. It has proven useful, in particular, to ease the characterization of quantum protocols and dynamics~\cite{carrasquilla2019reconstructing, tairelly, Santagatieaap9646, agresti2019pattern,lumino2018,rocchetto2019,butler2018,fischer2006,melnikov,wangpaesani,Cimini2020}.
In the context of structured light, \acp{NN} have been used to classify \ac{OAM} states of classical light for long distance free-space communication, even in the presence of environmental turbulence~\cite{krenn_2014,krenn_2016,Doster_17, Park_18, Lohani_turbo_18, Li_18}.
In this Letter, we apply \ac{ML} to characterize experimental \acp{VVB} generated using a platform
based on photonic {\acp{QW}} in the \ac{OAM} and polarization degrees of freedom~\cite{Innocenti2017,giordani_2018}.
Our approach requires neither additional interferometry stabilization nor spatial filtering, thus providing a robust strategy to decode information stored in \acp{VVB}, and is therefore a promising pathway towards managing higher-dimensional quantum systems. 

 We leverage both supervised and unsupervised learning techniques. We start by training a \ac{CNN} to classify experimental images belonging to predefined classes of states. This method gives good prediction accuracy, while remaining fairly problem-agnostic and thus useful for diverse applications. However, while providing high prediction accuracy, NN-based methods are difficult to interpret.
We thus also propose an alternative technique based on the joint application of {\ac{DR}} and supervised learning.
This method provides a geometrical description
of the underlying space associated to the experimental data.
While significantly easier to use, such approach gives comparable results to CNN,
at the cost of being more tailored to the specifics of the problem.

Our work makes significant steps forward  with respect to previous endeavours: while Refs.~\cite{krenn_2014,krenn_2016,Doster_17, Park_18, Lohani_turbo_18, Li_18} leverage NNs to process OAM states, our work is the first to tackle VVBs. Moreover, owing to the variety of techniques we deploy, we can address both classification and regression tasks, thus enabling the reconstruction of the input states in relevant cases of structured light beams.
Our findings demonstrate the reliability of a broader class of ML methods, providing novel recognition methods to deal with VVB, which are a building block for several information protocols with high-dimensional systems.

\begin{figure}[t]
    \includegraphics[width=0.5\textwidth]{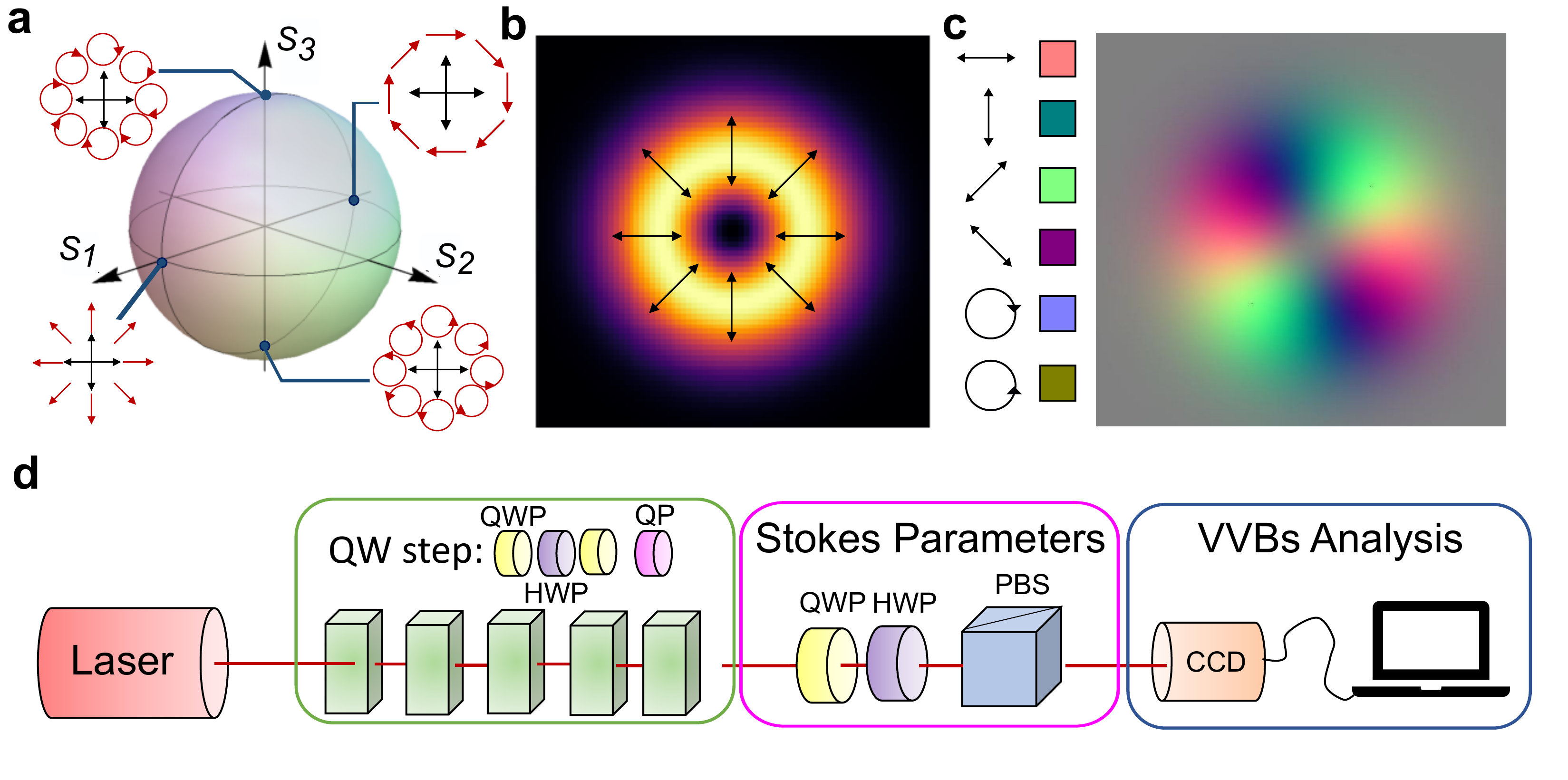}
    \caption{{\bf a,} Higher-order Poincar\'e sphere representation for $|m_{1,2}|=1$. Each point on the sphere surface corresponds to specific polarization patterns. 
    {\bf b,} A radially polarized \ac{VVB}: at a given point in the transverse plane the polarization vector has a different orientation. The Stokes parameters vary accordingly in the plane.
    {\bf c,} Color encoding of the polarization pattern. 
    The legend reports the correspondence between colors and the various polarizations.
    On the right we have the resulting color pattern for the VVB in panel {\bf b}.
    Grey color corresponds to unpolarized light.
    {\bf d,} Experimental apparatus for the generation of \ac{VVB}s. A continuous-wave laser emits a Gaussian beam ${\rm TEM}_{00}$ at $808$ nm. Light undergoes a 5-step quantum walk realized through a sequence of waveplates and q-plates.
    A CCD camera-based detection stage acquires information on the Stokes parameters and the polarization pattern. Based on the intensity measured at each pixels of the  camera, Stokes parameters are evaluated and converted into RGB-colored pictures.
    }%
    \label{poinc_sphere}
\end{figure}

\textit{Experimental generation of Vector Vortex Beams--} 
OAM-endowed states of light 
can be described using {\ac{LG}} modes.
These are solutions of the Helmholtz equation in the paraxial approximation, indexed by two integer numbers $(m, p)$, the former describing the azimuthal phase structure of the beam, and the latter describing its radial intensity profile.
Each~\ac{LG} mode carries a set amount of angular momentum, which in the single-photon regime equals $\hbar m$~\cite{allen_0AM_1992}.
\acp{VVB} can be obtained by superposing orthogonal polarizations to  LG modes ~\cite{padgett2004light}.
More specifically, the electric field $\Vec{E}_{m_1m_2p}$ of a \ac{VVB}  decomposes as the sum of two~\ac{LG} modes with same $p$ and different azimuthal numbers $m_1>m_2$ carried by orthogonal polarizations:
$\Vec{E}_{m_1m_2p}=\Vec{e}_L \cos{ \frac{\theta}{2}}\text{ LG$_{m_1p}$} +\Vec{e}_R e^{i \phi} \sin{ \frac{\theta}{2}}\text{ LG$_{m_2p}$}$,
where $\theta\in[0,\pi], \phi\in[0,2\pi]$ and the unit vectors $\Vec{e}_{L,R}$ stand for left and right circular polarization, respectively.
For the purpose of this work we can ignore the radial number, setting $p=0$.
For any given value of the parameters $(m_1$, $m_2, \theta, \phi)$,  the polarization pattern of a \ac{VVB} can be mapped onto a generalized Poincar\'e sphere (cf. Fig.~\ref{poinc_sphere}). In particular, we use the higher-order Poincar\'e representation in which the poles represent eigenstates of the total angular momentum but with opposite signs~\cite{milione_poinc_sphere_2011}.
These polarization patterns are reconstructed via the \emph{Stokes parameters} $S_{j}~(j=1,2,3)$, obtained by
measuring the output intensities $I_{b_j,1},I_{b_j,2}$ associated to a given choice of polarization basis $\{b_j \}=\{b_1=( H,V )$, $b_2=( D,A )$, $b_3=( L,R )\}$ as $S_{b_j}=(I_{b_j,1}-I_{b_j,2})/(I_{b_j,1}+I_{b_j,2})$. 
For a VVB, the values of $S_j$ depend on the coordinates in the transverse propagation plane {\cite{Cardano:12}}.
To visualize the polarization patterns of \acp{VVB}, we use an RGB color encoding in which the values of $S_j$ are interpreted as strengths of the corresponding color. In Fig.~\ref{poinc_sphere}{\bf b} and {\bf c} we report an example of such color-map for radially polarized \acp{VVB}.
A natural way to generate~\acp{VVB} is using \emph{q-plates}~\cite{marrucci2006optical,Cardano:12}, which are inhomogenous birefringent plates modifying the OAM of the incoming light conditionally to its polarization. 
In our scheme,
\acp{VVB} are generated via a sequence of polarization-controlling waveplates interspersing 5 cascaded q-plates (cf. Fig.~\ref{poinc_sphere}{\bf d}).
The apparatus implements a discrete-time QW in the angular momentum, where the order of LG modes takes the role of the \emph{walker} and it is changed according to the polarization state, which embodies the \emph{coin} degree of freedom~\cite{zhang-oam-qw-2010,goyal2013implementing,cardano2015quantum,Innocenti2017,giordani_2018}.
This allows to generate several classes of VVBs with OAM quantum numbers taking odd values in the interval $\{-5,..,5\}$.
We then collect images associated with different \acp{VVB} and use them to train and benchmark our ML-based approaches to classification , as discussed in the next sections.

\begin{figure}[b]
    \centering
     \includegraphics[width=\columnwidth]{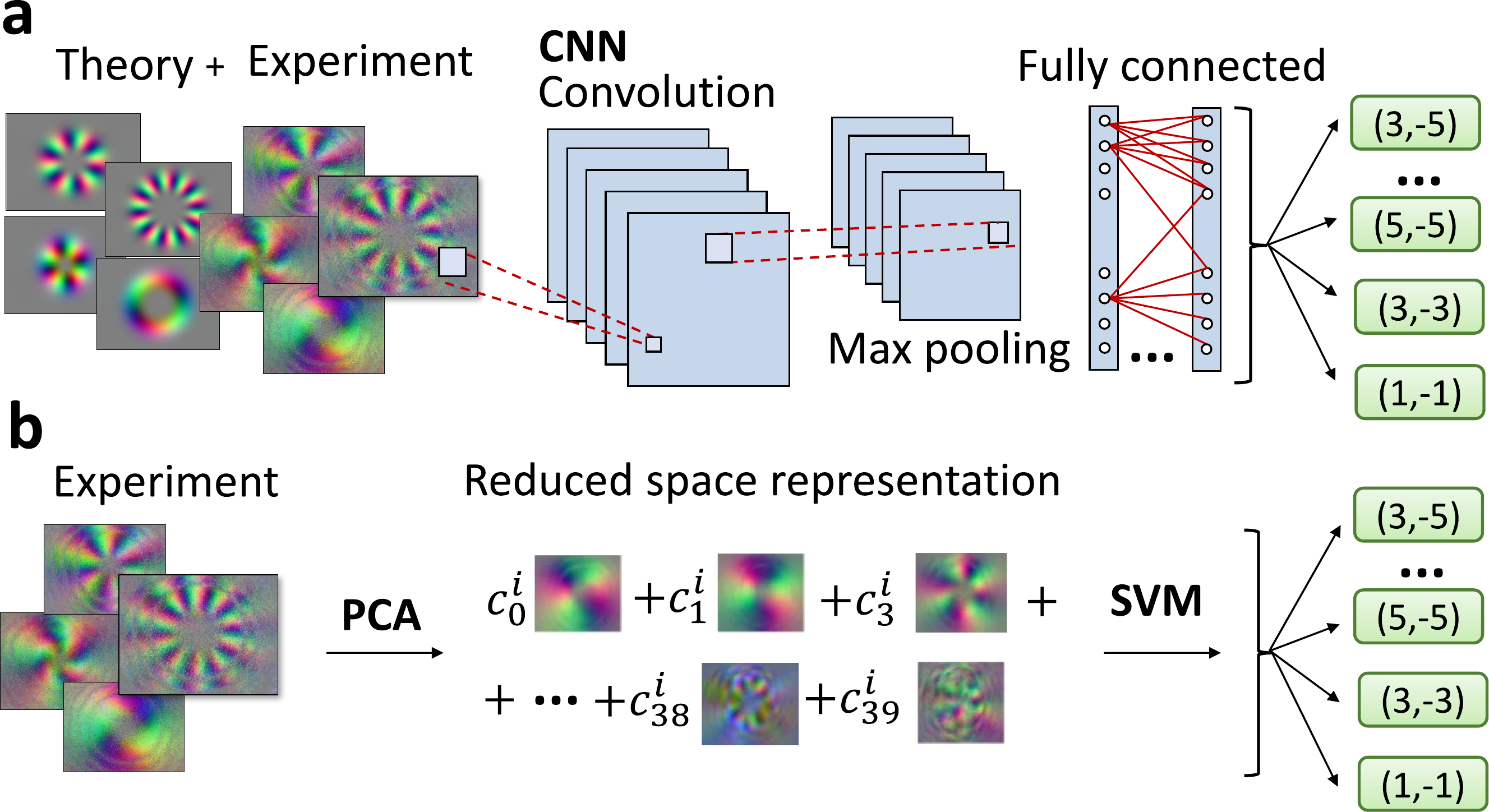}
    \caption{
    {\bf a,} 
Schematic representation of VVBs classification via CNNs.
    {\bf b,} 
    Classification scheme using linear PCA.
    After reducing the dimensionality of the dataset via PCA, a linear SVM is used to classify experimental images.} 
  
    \label{fig:class_techniques}
\end{figure}

\begin{figure}
    \centering
    \includegraphics[width=\columnwidth]{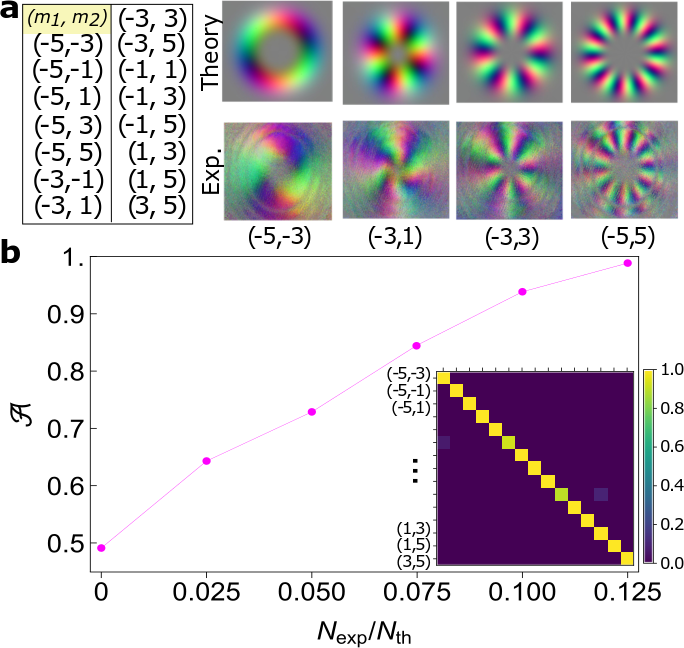}
    \caption{ 
    \textbf{a,} Simulated and experimental images of \acp{VVB} corresponding to some of the values $(m_1,m_2)$ given in the table.
    \textbf{b,} Scaling of the average accuracy $\mathcal{A}$ when classifying states into one of the $15$ VVB classes, against the fraction of experimental images added to the training set. The leftmost point refers to the case in which only simulated images are used to train the network.
    Inset: truth table reporting how the network classifies images belonging to each class.
    Each row (column) corresponds to a possible pair $(m_1,m_2)$.
    The matrix elements have been averaged over $100$ experimental images per class. 
    }
    \label{fig:resultsCNN}
\end{figure}

\textit{Classification via Convolutional Neural Networks--} We show here how to train a \ac{CNN}  to retrieve the parameters $(m_1,m_2)$ characterizing a given VVB from experimentally measured Stokes parameters.
\acp{CNN} are translation-invariant deep NNs well-suited for image classification ~\cite{lecun2015deep}, to recognize off-center images and segmented handwritten digits~\cite{Simard2003,Ciresan:2011:FHP:2283516.2283603},
and for facial recognition tasks~\cite{MATSUGU2003555}. 
In their simplest form, \acp{CNN} work by first applying a \emph{convolutional layer}, which consists of a series of nonlinear transformations applied to the input images, followed by a \emph{max-pooling layer}, which downsamples and filters the information extracted by the previous layer. Finally, a fully connected layer operates as a \emph{classifier}, categorizing the information extracted in the previous layers into one of a small number of possible output categories
(cf. \cite{SI,github,chollet2015keras,tensorflow2015-whitepaper,ruder2016overview} and Fig.~\ref{fig:class_techniques}).
 
The network is first fed with a training set made out of simulated images of \ac{VVB}s achievable with a five-step \ac{QW}.
The task is then to discern between $15$ classes, corresponding to the pairs $(m_1,m_2)$ in Fig.~\ref{fig:resultsCNN}{\bf a}.
For each class we generate states with $\theta=\pi/2$ and $\phi\in[0,2\pi]$. The size of the training set is $400$ images per class. Additional $100$ simulated images per class are used to benchmark the performance during training. 
In these conditions, the network achieves an accuracy of $100\%$.
The term \textit{accuracy} is used here to refer to the fraction of correctly classified images.
We then collect $100$ experimental images per class, to use as new validation set (cf. Fig.~\ref{fig:class_techniques}{\bf a}).
Fig.~\ref{fig:resultsCNN}{\bf a}-{\bf b} shows the average accuracy per class against the fraction of experimental images added to the training set.
The addition of a small fraction of experimental images to the training set improves the capability of the network to take into account deviations of the experimental states from ideal LG modes \cite{karimi_07,karimi_09, 
Rafayelyan,Shu:16,Vallone:16}
(cf. Fig.~\ref{fig:resultsCNN}{\bf b}). An average accuracy of $\sim 0.989$ is already obtained when $12.5\%$ of the training set is composed of experimental images.
To further highlight the performance of the network, 
we also trained a CNN using exclusively experimental images, but using a small number of images in the training phase. Using only $20$ images per class, we already get an accuracy of $0.99$ to classify the rest of the experimental images (which are $1668$ in total).

We use a similar approach to retrieve the position on the Poincar\'e sphere corresponding to states generated with fixed $(m_1,m_2)$.
In particular, we test the performance of CNNs to retrieve the values $(\theta,\phi)$ of VVBs corresponding to $m_2=-m_1=1$.
The CNN is thus trained to discriminate both rotations in the polarization patterns (corresponding to changes of $\phi$), and variations in the color tone (corresponding to changes of $\theta$).
To frame this as a classification task, we partition the sphere in $26$ disjoint sectors.
Working in spherical coordinates, we partition
$\theta$ in $3$ intervals $\left[k \frac{\pi}{8}, (k+2) \frac{\pi}{8}\right]$ with $k=1,3,5$, and $\phi$ in the $8$ intervals $\left[t \frac{\pi}{4}, (t+1) \frac{\pi}{4}\right]$ with $t \in \{0,...,7\} $.
This leaves two classes, surrounding the two poles, corresponding to $\theta \in \left[0, \frac{\pi}{8}\right]$ and $ \theta\in\left[ \frac{7}{8} \pi, \pi\right]$.
We train the CNN with $500$ simulated images per class in the training set, and $125$ per class in the validation one. The maximum achieved accuracy is $\sim 0.90$.
The sub-optimality of this result is likely a consequence of framing the problem as a classification task. Indeed, partitioning makes VVBs close to the border of two sectors naturally hard to classify. Training a CNN for the corresponding regression task will potentially improve performance.

\emph{Dimensionality reduction--}
We now present an alternative approach to classify {\acp{VVB}} from experimental data, 
leveraging  Dimensionality Reduction ({\ac{DR})}.
Such algorithms are typically used to obtain efficient representations of large datasets~\cite{cunningham2008dimension,fodor2002survey}.
This has several advantages, from easing data visualisation, to improving the efficiency of classification and regression algorithms, which can be used on the reduced representation of the data.
In particular, we employ a linear 
{\ac{PCA} algorithm, which} works by representing each datapoint as a vector in some high-dimensional space $\mathbb R^n$, and finding the directions in such space that capture the maximum amount of information about the dataset~\cite{jolliffe2011principal,jolliffe2016principal}.
The rationale for using {\ac{PCA}} in this context is that, although experimental images live in extremely high-dimensional spaces (whose dimension is of the order of the number of pixels in the CCD camera), the underlying dimension of the generated {\acp{VVB}} is typically much lower.
This means that, although the experimental dataset will \emph{a priori} seem like a complicated bundle of high-dimensional vectors, the underlying data is actually characterizable by a small number of parameters. Furthermore, the linearity of the mapping preserves the convexity of the VVB space and thus its geometrical structure. We then expect that the new description for expressing the experimental images in the reduced space provides a synthetic description for capturing the features of VVBs encoded in the measurements (the intensities in three polarization bases $\{b_j\}$, cf. \mbox{\cite{SI}}).
This resembles a form of \emph{unsupervised} learning, as we gain useful information about the origin of the images without feeding the algorithm with any knowledge of the underlying process.
 
\begin{figure}[t]
    \centering
    \includegraphics[width=\columnwidth]{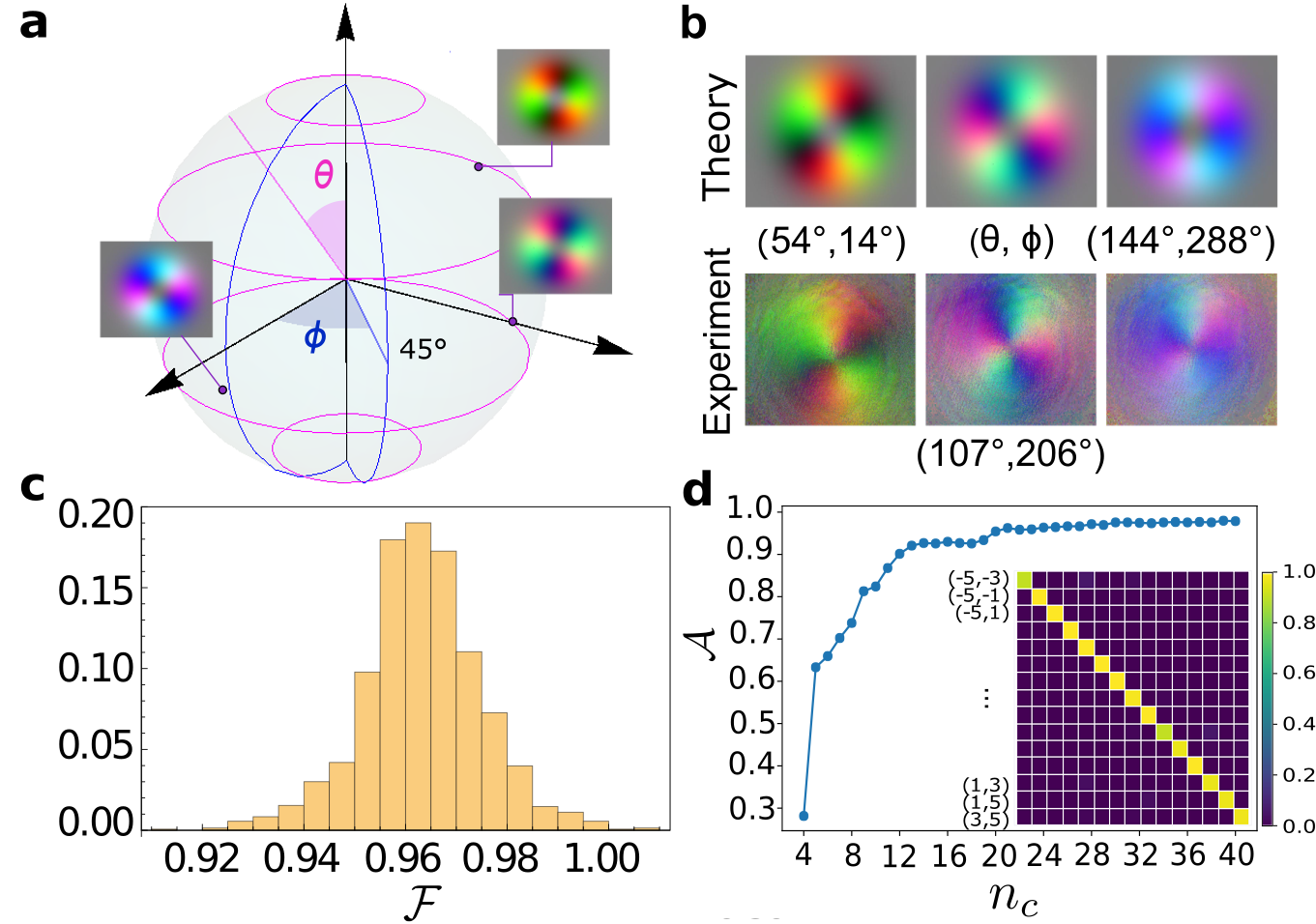}
    \caption{
    \textbf{a,}
    Higher-order Poincar\'e sphere for \ac{VVB}s with $|m_{1,2}|=1$. Magenta-colored parallels (Blue-colored meridians) mark intervals between consecutive values of $\theta$ ($\phi$). 
    Along a meridian the colors of the pattern vary from the hottest to the coldest one. Along a parallel, the patterns rotate. 
    \textbf{b,}
    Comparison between experimental and simulated \ac{VVB} images for different angles $(\theta, \phi)$.
    \textbf{c,}
    Distribution of fidelities obtained comparing each experimental VVBs with its reduced 3D representations given by PCA.
    \textbf{d,}
    Average prediction accuracy $\mathcal{A}$ of a linear SVM classifier, trained and tested after applying linear DR to the data, against the number of reduced dimensions $n_c$.
    For each of the 15 classes (cf. Fig.~\ref{fig:resultsCNN}a) in which the experimental dataset was divided, we show in the inset the truth table. 
    }%
    \label{fig:PCAresults}
\end{figure}

As a notable example, we apply these observations to \acp{VVB} with $m_2=-m_1=1$, which can be represented on a sphere in the higher-order Poincar\'e representation. 
Indeed, applying PCA to the experimental dataset of Fig.~\ref{fig:PCAresults}{\bf b}, reveals that three directions are sufficient to capture most of the information content of the images. Projecting the images along these three principal components, we find that the data are arranged in the form of a three-dimensional sphere embedded in the  experimental high-dimensional space. We refer to the supplementary material for the distribution of radii of the three-dimensional representation of the images \cite{SI} that allows to retrieve the state's position on the Poincar\'e sphere overcoming the border problem characterizing the previous classification method.
Remarkably, this was not obvious from the experimental dataset alone, but was easily revealed using \ac{DR}. This result highlights the potential of \ac{DR} to reveal features of the underlying states generating a given experimental dataset in realistic experimental conditions (cf.~\cite{SI,github}).
Interpreting this reduced three-dimensional representation as a Bloch sphere, we can use PCA to retrieve a complete description of the state generating a given experimental image.
To assess the accuracy of such reconstruction, we compute the average fidelity $\mathcal F_{\text{avg}}$ between the state generating a given image and the one retrieved from said image via PCA, averaging over many experimental images.
The fidelity between two states is here defined in the usual way as $\mathcal F(\rho,\sigma)\equiv \Tr|\sqrt\rho\sqrt\sigma|$.
As shown in the histogram of Fig.~\ref{fig:PCAresults}c, this is found to be $\mathcal F_{\text{avg}}\sim0.96$, with standard deviation $\sim0.01$, thus showcasing the quality of the reconstruction.

\textit{Classification via SVMs}-- We now show how the reduced representations provided by \ac{PCA} can function as starting point to train a classifier with accuracy comparable with the \ac{CNN}, whilst requiring a significantly reduced amount of computational resources.
More precisely, we use as classifiers linear \acp{SVM}~\cite{Hearst_si,cristianini_shawe-taylor_2000}. These supervised learning algorithms categorize data by finding the hyperplane that optimally separates the training dataset in accordance with the corresponding labels. 

As done for the CNN, we consider the task of classifying experimental dataset of \ac{VVB} states, indexed by $(m_1,m_2)$. 
We train the \ac{SVM} on the reduced space obtained via {\ac{PCA}}, applied to the experimental dataset reported in Fig.~\ref{fig:resultsCNN}{\bf a}. This significantly improves the efficiency of the classifier, which only has to operate on a compressed representation of the images.
This method gives an average accuracy of $\sim98\%$ when reducing the dimensionality of the dataset to $40$ \cite{SI,github}.
The~\ac{SVM} was trained on half of the experimental data, with the other half used to test the resulting accuracy. A breakdown of the resulting classification performance is reported in the inset of Fig.~\ref{fig:PCAresults}{\bf d}, in which we give the accuracy of the classifier for each class.
Finally, we highlight in Fig.~\ref{fig:PCAresults}{\bf d} how the average overall accuracy depends on the dimensionality of the reduced representation. In particular, we find that $\sim 25$ dimensions are already sufficient to get good average accuracies.

\textit{Discussion--} We presented a novel approach to classify \acp{VVB} leveraging ML techniques. We demonstrated how the use of inference strategies based on CNNs and PCA (enhanced by SVMs) allows to extract efficiently properties of high-dimensional photonic {\ac{VVB}} systems.
In particular, DR was used to obtain a deeper understanding of the underlying geometrical properties of the experimentally generated states, without requiring prior knowledge about the physics of the generation apparatus.
By embedding a variety of {\ac{ML}} algorithms into our experimental pipeline, the task of characterising structured light is made significantly broader in the methods, ranging from supervised to unsupervised learning, and more flexible in the applications, classification and regression tasks.
While paving the way to further experimental validations -- potentially also in experimental settings that do not rely on optical networks -- we believe that numerous tasks of relevance to modern photonics could benefit from introducing similar {\ac{ML}} ideas into their characterization protocols. These techniques can prove to be useful add-on to tasks ranging from the design of automatized approaches to the characterization of experimental platforms and experiments, to the provision of solutions to OAM demultiplexing in the context of classical and quantum communication and, more generally, for the use of structured light in quantum technologies.

\textit{Note--} During the reviewing process of this manuscript, the authors became aware of a related work~\cite{Zhanwei_oamCNN}, that addresses the classification of scalar fields with fractional topological charge. 

\begin{acknowledgments}
\textit{Acknowledgments--} We acknowledge support from the ERC Advanced grant PHOSPhOR (Photonics of Spin-Orbit Optical Phenomena; 
Grant Agreement No. 828978), the EU Collaborative project TEQ (grant nr. 766900), Fondazione Angelo della Riccia, the DfE-SFI Investigator Programme (grant 15/IA/2864), COST Action CA15220, the Royal Society Wolfson Research Fellowship (RSWF\textbackslash R3\textbackslash183013), the Leverhulme Trust Research Project Grant (grant nr.~RGP-2018-266).
\end{acknowledgments}


\begin{thebibliography}{82}%
\makeatletter
\providecommand \@ifxundefined [1]{%
 \@ifx{#1\undefined}
}%
\providecommand \@ifnum [1]{%
 \ifnum #1\expandafter \@firstoftwo
 \else \expandafter \@secondoftwo
 \fi
}%
\providecommand \@ifx [1]{%
 \ifx #1\expandafter \@firstoftwo
 \else \expandafter \@secondoftwo
 \fi
}%
\providecommand \natexlab [1]{#1}%
\providecommand \enquote  [1]{``#1''}%
\providecommand \bibnamefont  [1]{#1}%
\providecommand \bibfnamefont [1]{#1}%
\providecommand \citenamefont [1]{#1}%
\providecommand \href@noop [0]{\@secondoftwo}%
\providecommand \href [0]{\begingroup \@sanitize@url \@href}%
\providecommand \@href[1]{\@@startlink{#1}\@@href}%
\providecommand \@@href[1]{\endgroup#1\@@endlink}%
\providecommand \@sanitize@url [0]{\catcode `\\12\catcode `\$12\catcode
  `\&12\catcode `\#12\catcode `\^12\catcode `\_12\catcode `\%12\relax}%
\providecommand \@@startlink[1]{}%
\providecommand \@@endlink[0]{}%
\providecommand \url  [0]{\begingroup\@sanitize@url \@url }%
\providecommand \@url [1]{\endgroup\@href {#1}{\urlprefix }}%
\providecommand \urlprefix  [0]{URL }%
\providecommand \Eprint [0]{\href }%
\providecommand \doibase [0]{http://dx.doi.org/}%
\providecommand \selectlanguage [0]{\@gobble}%
\providecommand \bibinfo  [0]{\@secondoftwo}%
\providecommand \bibfield  [0]{\@secondoftwo}%
\providecommand \translation [1]{[#1]}%
\providecommand \BibitemOpen [0]{}%
\providecommand \bibitemStop [0]{}%
\providecommand \bibitemNoStop [0]{.\EOS\space}%
\providecommand \EOS [0]{\spacefactor3000\relax}%
\providecommand \BibitemShut  [1]{\csname bibitem#1\endcsname}%
\let\auto@bib@innerbib\@empty
\bibitem [{\citenamefont {Allen}\ \emph {et~al.}(1992)\citenamefont {Allen},
  \citenamefont {Beijersbergen}, \citenamefont {Spreeuw},\ and\ \citenamefont
  {Woerdman}}]{allen_0AM_1992}%
  \BibitemOpen
  \bibfield  {author} {\bibinfo {author} {\bibfnamefont {L.}~\bibnamefont
  {Allen}}, \bibinfo {author} {\bibfnamefont {M.~W.}\ \bibnamefont
  {Beijersbergen}}, \bibinfo {author} {\bibfnamefont {R.~J.~C.}\ \bibnamefont
  {Spreeuw}}, \ and\ \bibinfo {author} {\bibfnamefont {J.~P.}\ \bibnamefont
  {Woerdman}},\ }\href {\doibase 10.1103/PhysRevA.45.8185} {\bibfield
  {journal} {\bibinfo  {journal} {Phys. Rev. A}\ }\textbf {\bibinfo {volume}
  {45}},\ \bibinfo {pages} {8185} (\bibinfo {year} {1992})}\BibitemShut
  {NoStop}%
\bibitem [{\citenamefont {Padgett}\ \emph {et~al.}(2004)\citenamefont
  {Padgett}, \citenamefont {Courtial},\ and\ \citenamefont
  {Allen}}]{padgett2004light}%
  \BibitemOpen
  \bibfield  {author} {\bibinfo {author} {\bibfnamefont {M.}~\bibnamefont
  {Padgett}}, \bibinfo {author} {\bibfnamefont {J.}~\bibnamefont {Courtial}}, \
  and\ \bibinfo {author} {\bibfnamefont {L.}~\bibnamefont {Allen}},\ }\href
  {\doibase 10.1063/1.1768672} {\bibfield  {journal} {\bibinfo  {journal}
  {Physics Today}\ }\textbf {\bibinfo {volume} {57}},\ \bibinfo {pages} {35}
  (\bibinfo {year} {2004})}\BibitemShut {NoStop}%
\bibitem [{\citenamefont {Erhard}\ \emph {et~al.}(2018)\citenamefont {Erhard},
  \citenamefont {Fickler}, \citenamefont {Krenn},\ and\ \citenamefont
  {Zeilinger}}]{erhard2018twisted}%
  \BibitemOpen
  \bibfield  {author} {\bibinfo {author} {\bibfnamefont {M.}~\bibnamefont
  {Erhard}}, \bibinfo {author} {\bibfnamefont {R.}~\bibnamefont {Fickler}},
  \bibinfo {author} {\bibfnamefont {M.}~\bibnamefont {Krenn}}, \ and\ \bibinfo
  {author} {\bibfnamefont {A.}~\bibnamefont {Zeilinger}},\ }\href {\doibase
  10.1038/lsa.2017.146} {\bibfield  {journal} {\bibinfo  {journal} {Light:
  Science \& Applications}\ }\textbf {\bibinfo {volume} {7}},\ \bibinfo {pages}
  {17146} (\bibinfo {year} {2018})}\BibitemShut {NoStop}%
\bibitem [{\citenamefont {Marrucci}\ \emph {et~al.}(2011)\citenamefont
  {Marrucci}, \citenamefont {Karimi}, \citenamefont {Slussarenko},
  \citenamefont {Piccirillo}, \citenamefont {Santamato}, \citenamefont
  {Nagali},\ and\ \citenamefont {Sciarrino}}]{Marrucci2011Rev}%
  \BibitemOpen
  \bibfield  {author} {\bibinfo {author} {\bibfnamefont {L.}~\bibnamefont
  {Marrucci}}, \bibinfo {author} {\bibfnamefont {E.}~\bibnamefont {Karimi}},
  \bibinfo {author} {\bibfnamefont {S.}~\bibnamefont {Slussarenko}}, \bibinfo
  {author} {\bibfnamefont {B.}~\bibnamefont {Piccirillo}}, \bibinfo {author}
  {\bibfnamefont {E.}~\bibnamefont {Santamato}}, \bibinfo {author}
  {\bibfnamefont {E.}~\bibnamefont {Nagali}}, \ and\ \bibinfo {author}
  {\bibfnamefont {F.}~\bibnamefont {Sciarrino}},\ }\href {\doibase
  10.1088/2040-8978/13/6/064001} {\bibfield  {journal} {\bibinfo  {journal} {J.
  Opt}\ }\textbf {\bibinfo {volume} {13}},\ \bibinfo {pages} {064001} (\bibinfo
  {year} {2011})}\BibitemShut {NoStop}%
\bibitem [{\citenamefont {Cozzolino}\ \emph
  {et~al.}(2019{\natexlab{a}})\citenamefont {Cozzolino}, \citenamefont
  {Da~Lio}, \citenamefont {Bacco},\ and\ \citenamefont
  {Oxenløwe}}]{Cozzolino_rev}%
  \BibitemOpen
  \bibfield  {author} {\bibinfo {author} {\bibfnamefont {D.}~\bibnamefont
  {Cozzolino}}, \bibinfo {author} {\bibfnamefont {B.}~\bibnamefont {Da~Lio}},
  \bibinfo {author} {\bibfnamefont {D.}~\bibnamefont {Bacco}}, \ and\ \bibinfo
  {author} {\bibfnamefont {L.~K.}\ \bibnamefont {Oxenløwe}},\ }\href {\doibase
  10.1002/qute.201900038} {\bibfield  {journal} {\bibinfo  {journal} {Advanced
  Quantum Technologies}\ }\textbf {\bibinfo {volume} {2}},\ \bibinfo {pages}
  {1900038} (\bibinfo {year} {2019}{\natexlab{a}})},\ \Eprint
  {http://arxiv.org/abs/https://onlinelibrary.wiley.com/doi/pdf/10.1002/qute.201900038}
  {https://onlinelibrary.wiley.com/doi/pdf/10.1002/qute.201900038} \BibitemShut
  {NoStop}%
\bibitem [{\citenamefont {Cardano}\ and\ \citenamefont
  {Marrucci}(2015)}]{Cardano2015Rev}%
  \BibitemOpen
  \bibfield  {author} {\bibinfo {author} {\bibfnamefont {F.}~\bibnamefont
  {Cardano}}\ and\ \bibinfo {author} {\bibfnamefont {L.}~\bibnamefont
  {Marrucci}},\ }\href {https://doi.org/10.1038/nphoton.2015.232} {\bibfield
  {journal} {\bibinfo  {journal} {Nature Photonics}\ }\textbf {\bibinfo
  {volume} {9}},\ \bibinfo {pages} {776} (\bibinfo {year} {2015})}\BibitemShut
  {NoStop}%
\bibitem [{\citenamefont {Rubinsztein-Dunlop}\ \emph
  {et~al.}(2017)\citenamefont {Rubinsztein-Dunlop}, \citenamefont {Forbes},
  \citenamefont {Berry}, \citenamefont {Dennis}, \citenamefont {Andrews},
  \citenamefont {Mansuripur}, \citenamefont {Denz}, \citenamefont {Alpmann},
  \citenamefont {Banzer}, \citenamefont {Bauer}, \citenamefont {Karimi},
  \citenamefont {Marrucci}, \citenamefont {Padgett}, \citenamefont
  {Ritsch-Marte}, \citenamefont {Litchinitser}, \citenamefont {Bigelow},
  \citenamefont {Rosales-Guzmán}, \citenamefont {Belmonte}, \citenamefont
  {Torres}, \citenamefont {Neely}, \citenamefont {Baker}, \citenamefont
  {Gordon}, \citenamefont {Stilgoe}, \citenamefont {Romero}, \citenamefont
  {White}, \citenamefont {Fickler}, \citenamefont {Willner}, \citenamefont
  {Xie}, \citenamefont {McMorran},\ and\ \citenamefont {Weiner}}]{roadMap}%
  \BibitemOpen
  \bibfield  {author} {\bibinfo {author} {\bibfnamefont {H.}~\bibnamefont
  {Rubinsztein-Dunlop}}, \bibinfo {author} {\bibfnamefont {A.}~\bibnamefont
  {Forbes}}, \bibinfo {author} {\bibfnamefont {M.~V.}\ \bibnamefont {Berry}},
  \bibinfo {author} {\bibfnamefont {M.~R.}\ \bibnamefont {Dennis}}, \bibinfo
  {author} {\bibfnamefont {D.~L.}\ \bibnamefont {Andrews}}, \bibinfo {author}
  {\bibfnamefont {M.}~\bibnamefont {Mansuripur}}, \bibinfo {author}
  {\bibfnamefont {C.}~\bibnamefont {Denz}}, \bibinfo {author} {\bibfnamefont
  {C.}~\bibnamefont {Alpmann}}, \bibinfo {author} {\bibfnamefont
  {P.}~\bibnamefont {Banzer}}, \bibinfo {author} {\bibfnamefont
  {T.}~\bibnamefont {Bauer}}, \bibinfo {author} {\bibfnamefont
  {E.}~\bibnamefont {Karimi}}, \bibinfo {author} {\bibfnamefont
  {L.}~\bibnamefont {Marrucci}}, \bibinfo {author} {\bibfnamefont
  {M.}~\bibnamefont {Padgett}}, \bibinfo {author} {\bibfnamefont
  {M.}~\bibnamefont {Ritsch-Marte}}, \bibinfo {author} {\bibfnamefont {N.~M.}\
  \bibnamefont {Litchinitser}}, \bibinfo {author} {\bibfnamefont {N.~P.}\
  \bibnamefont {Bigelow}}, \bibinfo {author} {\bibfnamefont {C.}~\bibnamefont
  {Rosales-Guzmán}}, \bibinfo {author} {\bibfnamefont {A.}~\bibnamefont
  {Belmonte}}, \bibinfo {author} {\bibfnamefont {J.~P.}\ \bibnamefont
  {Torres}}, \bibinfo {author} {\bibfnamefont {T.~W.}\ \bibnamefont {Neely}},
  \bibinfo {author} {\bibfnamefont {M.}~\bibnamefont {Baker}}, \bibinfo
  {author} {\bibfnamefont {R.}~\bibnamefont {Gordon}}, \bibinfo {author}
  {\bibfnamefont {A.~B.}\ \bibnamefont {Stilgoe}}, \bibinfo {author}
  {\bibfnamefont {J.}~\bibnamefont {Romero}}, \bibinfo {author} {\bibfnamefont
  {A.~G.}\ \bibnamefont {White}}, \bibinfo {author} {\bibfnamefont
  {R.}~\bibnamefont {Fickler}}, \bibinfo {author} {\bibfnamefont {A.~E.}\
  \bibnamefont {Willner}}, \bibinfo {author} {\bibfnamefont {G.}~\bibnamefont
  {Xie}}, \bibinfo {author} {\bibfnamefont {B.}~\bibnamefont {McMorran}}, \
  and\ \bibinfo {author} {\bibfnamefont {A.~M.}\ \bibnamefont {Weiner}},\
  }\href {\doibase 10.1088/2040-8978/19/1/013001} {\bibfield  {journal}
  {\bibinfo  {journal} {J. Opt.}\ }\textbf {\bibinfo {volume} {19}},\ \bibinfo
  {pages} {013001} (\bibinfo {year} {2017})}\BibitemShut {NoStop}%
\bibitem [{\citenamefont {Willner}\ \emph {et~al.}(2015)\citenamefont
  {Willner}, \citenamefont {Huang}, \citenamefont {Yan}, \citenamefont {Ren},
  \citenamefont {Ahmed}, \citenamefont {Xie}, \citenamefont {Bao},
  \citenamefont {Li}, \citenamefont {Cao}, \citenamefont {Zhao}, \citenamefont
  {Wang}, \citenamefont {Lavery}, \citenamefont {Tur}, \citenamefont
  {Ramachandran}, \citenamefont {Molisch}, \citenamefont {Ashrafi},\ and\
  \citenamefont {Ashrafi}}]{Willner:15}%
  \BibitemOpen
  \bibfield  {author} {\bibinfo {author} {\bibfnamefont {A.~E.}\ \bibnamefont
  {Willner}}, \bibinfo {author} {\bibfnamefont {H.}~\bibnamefont {Huang}},
  \bibinfo {author} {\bibfnamefont {Y.}~\bibnamefont {Yan}}, \bibinfo {author}
  {\bibfnamefont {Y.}~\bibnamefont {Ren}}, \bibinfo {author} {\bibfnamefont
  {N.}~\bibnamefont {Ahmed}}, \bibinfo {author} {\bibfnamefont
  {G.}~\bibnamefont {Xie}}, \bibinfo {author} {\bibfnamefont {C.}~\bibnamefont
  {Bao}}, \bibinfo {author} {\bibfnamefont {L.}~\bibnamefont {Li}}, \bibinfo
  {author} {\bibfnamefont {Y.}~\bibnamefont {Cao}}, \bibinfo {author}
  {\bibfnamefont {Z.}~\bibnamefont {Zhao}}, \bibinfo {author} {\bibfnamefont
  {J.}~\bibnamefont {Wang}}, \bibinfo {author} {\bibfnamefont {M.~P.~J.}\
  \bibnamefont {Lavery}}, \bibinfo {author} {\bibfnamefont {M.}~\bibnamefont
  {Tur}}, \bibinfo {author} {\bibfnamefont {S.}~\bibnamefont {Ramachandran}},
  \bibinfo {author} {\bibfnamefont {A.~F.}\ \bibnamefont {Molisch}}, \bibinfo
  {author} {\bibfnamefont {N.}~\bibnamefont {Ashrafi}}, \ and\ \bibinfo
  {author} {\bibfnamefont {S.}~\bibnamefont {Ashrafi}},\ }\href {\doibase
  10.1364/AOP.7.000066} {\bibfield  {journal} {\bibinfo  {journal} {Adv. Opt.
  Photon.}\ }\textbf {\bibinfo {volume} {7}},\ \bibinfo {pages} {66} (\bibinfo
  {year} {2015})}\BibitemShut {NoStop}%
\bibitem [{\citenamefont {Cozzolino}\ \emph
  {et~al.}(2019{\natexlab{b}})\citenamefont {Cozzolino}, \citenamefont
  {Polino}, \citenamefont {Valeri}, \citenamefont {Carvacho}, \citenamefont
  {Bacco}, \citenamefont {Spagnolo}, \citenamefont {Oxenl{\o}we},\ and\
  \citenamefont {Sciarrino}}]{cozzolino2019air}%
  \BibitemOpen
  \bibfield  {author} {\bibinfo {author} {\bibfnamefont {D.}~\bibnamefont
  {Cozzolino}}, \bibinfo {author} {\bibfnamefont {E.}~\bibnamefont {Polino}},
  \bibinfo {author} {\bibfnamefont {M.}~\bibnamefont {Valeri}}, \bibinfo
  {author} {\bibfnamefont {G.}~\bibnamefont {Carvacho}}, \bibinfo {author}
  {\bibfnamefont {D.}~\bibnamefont {Bacco}}, \bibinfo {author} {\bibfnamefont
  {N.}~\bibnamefont {Spagnolo}}, \bibinfo {author} {\bibfnamefont {L.~K.}\
  \bibnamefont {Oxenl{\o}we}}, \ and\ \bibinfo {author} {\bibfnamefont
  {F.}~\bibnamefont {Sciarrino}},\ }\href {\doibase 10.1117/1.AP.1.4.046005}
  {\bibfield  {journal} {\bibinfo  {journal} {Advanced Photonics}\ }\textbf
  {\bibinfo {volume} {1}},\ \bibinfo {pages} {046005} (\bibinfo {year}
  {2019}{\natexlab{b}})}\BibitemShut {NoStop}%
\bibitem [{\citenamefont {Fickler}\ \emph {et~al.}(2012)\citenamefont
  {Fickler}, \citenamefont {Lapkiewicz}, \citenamefont {Plick}, \citenamefont
  {Krenn}, \citenamefont {Schaeff}, \citenamefont {Ramelow},\ and\
  \citenamefont {Zeilinger}}]{fickler2012quantum}%
  \BibitemOpen
  \bibfield  {author} {\bibinfo {author} {\bibfnamefont {R.}~\bibnamefont
  {Fickler}}, \bibinfo {author} {\bibfnamefont {R.}~\bibnamefont {Lapkiewicz}},
  \bibinfo {author} {\bibfnamefont {W.~N.}\ \bibnamefont {Plick}}, \bibinfo
  {author} {\bibfnamefont {M.}~\bibnamefont {Krenn}}, \bibinfo {author}
  {\bibfnamefont {C.}~\bibnamefont {Schaeff}}, \bibinfo {author} {\bibfnamefont
  {S.}~\bibnamefont {Ramelow}}, \ and\ \bibinfo {author} {\bibfnamefont
  {A.}~\bibnamefont {Zeilinger}},\ }\href
  {https://doi.org/10.1126/science.1227193} {\bibfield  {journal} {\bibinfo
  {journal} {Science}\ }\textbf {\bibinfo {volume} {338}},\ \bibinfo {pages}
  {640} (\bibinfo {year} {2012})}\BibitemShut {NoStop}%
\bibitem [{\citenamefont {D'Ambrosio}\ \emph {et~al.}(2013)\citenamefont
  {D'Ambrosio}, \citenamefont {Spagnolo}, \citenamefont {{Del Re}},
  \citenamefont {Slussarenko}, \citenamefont {Li}, \citenamefont {Kwek},
  \citenamefont {Marrucci}, \citenamefont {Walborn}, \citenamefont {Aolita},\
  and\ \citenamefont {Sciarrino}}]{dambrosio_gear2013}%
  \BibitemOpen
  \bibfield  {author} {\bibinfo {author} {\bibfnamefont {V.}~\bibnamefont
  {D'Ambrosio}}, \bibinfo {author} {\bibfnamefont {N.}~\bibnamefont
  {Spagnolo}}, \bibinfo {author} {\bibfnamefont {L.}~\bibnamefont {{Del Re}}},
  \bibinfo {author} {\bibfnamefont {S.}~\bibnamefont {Slussarenko}}, \bibinfo
  {author} {\bibfnamefont {Y.}~\bibnamefont {Li}}, \bibinfo {author}
  {\bibfnamefont {L.~C.}\ \bibnamefont {Kwek}}, \bibinfo {author}
  {\bibfnamefont {L.}~\bibnamefont {Marrucci}}, \bibinfo {author}
  {\bibfnamefont {S.~P.}\ \bibnamefont {Walborn}}, \bibinfo {author}
  {\bibfnamefont {L.}~\bibnamefont {Aolita}}, \ and\ \bibinfo {author}
  {\bibfnamefont {F.}~\bibnamefont {Sciarrino}},\ }\href
  {https://www.nature.com/articles/ncomms3432} {\bibfield  {journal} {\bibinfo
  {journal} {Nat. Comm.}\ }\textbf {\bibinfo {volume} {4}},\ \bibinfo {pages}
  {2432} (\bibinfo {year} {2013})}\BibitemShut {NoStop}%
\bibitem [{\citenamefont {Goswami}\ \emph {et~al.}(2018)\citenamefont
  {Goswami}, \citenamefont {Giarmatzi}, \citenamefont {Kewming}, \citenamefont
  {Costa}, \citenamefont {Branciard}, \citenamefont {Romero},\ and\
  \citenamefont {White}}]{Goswami2018}%
  \BibitemOpen
  \bibfield  {author} {\bibinfo {author} {\bibfnamefont {K.}~\bibnamefont
  {Goswami}}, \bibinfo {author} {\bibfnamefont {C.}~\bibnamefont {Giarmatzi}},
  \bibinfo {author} {\bibfnamefont {M.}~\bibnamefont {Kewming}}, \bibinfo
  {author} {\bibfnamefont {F.}~\bibnamefont {Costa}}, \bibinfo {author}
  {\bibfnamefont {C.}~\bibnamefont {Branciard}}, \bibinfo {author}
  {\bibfnamefont {J.}~\bibnamefont {Romero}}, \ and\ \bibinfo {author}
  {\bibfnamefont {A.~G.}\ \bibnamefont {White}},\ }\href {\doibase
  10.1103/PhysRevLett.121.090503} {\bibfield  {journal} {\bibinfo  {journal}
  {Phys. Rev. Lett.}\ }\textbf {\bibinfo {volume} {121}},\ \bibinfo {pages}
  {090503} (\bibinfo {year} {2018})}\BibitemShut {NoStop}%
\bibitem [{\citenamefont {Vallone}\ \emph {et~al.}(2014)\citenamefont
  {Vallone}, \citenamefont {D’Ambrosio}, \citenamefont {Sponselli},
  \citenamefont {Slussarenko}, \citenamefont {Marrucci}, \citenamefont
  {Sciarrino},\ and\ \citenamefont {Villoresi}}]{vallone_qkd_2014}%
  \BibitemOpen
  \bibfield  {author} {\bibinfo {author} {\bibfnamefont {G.}~\bibnamefont
  {Vallone}}, \bibinfo {author} {\bibfnamefont {V.}~\bibnamefont
  {D’Ambrosio}}, \bibinfo {author} {\bibfnamefont {A.}~\bibnamefont
  {Sponselli}}, \bibinfo {author} {\bibfnamefont {S.}~\bibnamefont
  {Slussarenko}}, \bibinfo {author} {\bibfnamefont {L.}~\bibnamefont
  {Marrucci}}, \bibinfo {author} {\bibfnamefont {F.}~\bibnamefont {Sciarrino}},
  \ and\ \bibinfo {author} {\bibfnamefont {P.}~\bibnamefont {Villoresi}},\
  }\href {\doibase 10.1103/PhysRevLett.113.060503} {\bibfield  {journal}
  {\bibinfo  {journal} {Phys. Rev. Lett.}\ }\textbf {\bibinfo {volume} {113}},\
  \bibinfo {pages} {060503} (\bibinfo {year} {2014})}\BibitemShut {NoStop}%
\bibitem [{\citenamefont {Wang}\ \emph {et~al.}(2015)\citenamefont {Wang},
  \citenamefont {Cai}, \citenamefont {Su}, \citenamefont {Chen}, \citenamefont
  {Wu}, \citenamefont {Li}, \citenamefont {Liu}, \citenamefont {Lu},\ and\
  \citenamefont {Pan}}]{Wang2015}%
  \BibitemOpen
  \bibfield  {author} {\bibinfo {author} {\bibfnamefont {X.-L.}\ \bibnamefont
  {Wang}}, \bibinfo {author} {\bibfnamefont {X.-D.}\ \bibnamefont {Cai}},
  \bibinfo {author} {\bibfnamefont {Z.-E.}\ \bibnamefont {Su}}, \bibinfo
  {author} {\bibfnamefont {M.-C.}\ \bibnamefont {Chen}}, \bibinfo {author}
  {\bibfnamefont {D.}~\bibnamefont {Wu}}, \bibinfo {author} {\bibfnamefont
  {L.}~\bibnamefont {Li}}, \bibinfo {author} {\bibfnamefont {N.-L.}\
  \bibnamefont {Liu}}, \bibinfo {author} {\bibfnamefont {C.-Y.}\ \bibnamefont
  {Lu}}, \ and\ \bibinfo {author} {\bibfnamefont {J.-W.}\ \bibnamefont {Pan}},\
  }\href {\doibase 10.1038/nature14246} {\bibfield  {journal} {\bibinfo
  {journal} {Nature}\ }\textbf {\bibinfo {volume} {518}},\ \bibinfo {pages}
  {516} (\bibinfo {year} {2015})}\BibitemShut {NoStop}%
\bibitem [{\citenamefont {Mirhosseini}\ \emph {et~al.}(2015)\citenamefont
  {Mirhosseini}, \citenamefont {Maga{\~{n}}a-Loaiza}, \citenamefont
  {O'Sullivan}, \citenamefont {Rodenburg}, \citenamefont {Malik}, \citenamefont
  {Lavery}, \citenamefont {Padgett}, \citenamefont {Gauthier},\ and\
  \citenamefont {Boyd}}]{Mirhosseini_2015}%
  \BibitemOpen
  \bibfield  {author} {\bibinfo {author} {\bibfnamefont {M.}~\bibnamefont
  {Mirhosseini}}, \bibinfo {author} {\bibfnamefont {O.~S.}\ \bibnamefont
  {Maga{\~{n}}a-Loaiza}}, \bibinfo {author} {\bibfnamefont {M.~N.}\
  \bibnamefont {O'Sullivan}}, \bibinfo {author} {\bibfnamefont
  {B.}~\bibnamefont {Rodenburg}}, \bibinfo {author} {\bibfnamefont
  {M.}~\bibnamefont {Malik}}, \bibinfo {author} {\bibfnamefont {M.~P.~J.}\
  \bibnamefont {Lavery}}, \bibinfo {author} {\bibfnamefont {M.~J.}\
  \bibnamefont {Padgett}}, \bibinfo {author} {\bibfnamefont {D.~J.}\
  \bibnamefont {Gauthier}}, \ and\ \bibinfo {author} {\bibfnamefont {R.~W.}\
  \bibnamefont {Boyd}},\ }\href {\doibase 10.1088/1367-2630/17/3/033033}
  {\bibfield  {journal} {\bibinfo  {journal} {New Journal of Physics}\ }\textbf
  {\bibinfo {volume} {17}},\ \bibinfo {pages} {033033} (\bibinfo {year}
  {2015})}\BibitemShut {NoStop}%
\bibitem [{\citenamefont {Malik}\ \emph {et~al.}(2016)\citenamefont {Malik},
  \citenamefont {Erhard}, \citenamefont {Huber}, \citenamefont {Krenn},
  \citenamefont {Fickler},\ and\ \citenamefont {Zeilinger}}]{Malik2016}%
  \BibitemOpen
  \bibfield  {author} {\bibinfo {author} {\bibfnamefont {M.}~\bibnamefont
  {Malik}}, \bibinfo {author} {\bibfnamefont {M.}~\bibnamefont {Erhard}},
  \bibinfo {author} {\bibfnamefont {M.}~\bibnamefont {Huber}}, \bibinfo
  {author} {\bibfnamefont {M.}~\bibnamefont {Krenn}}, \bibinfo {author}
  {\bibfnamefont {R.}~\bibnamefont {Fickler}}, \ and\ \bibinfo {author}
  {\bibfnamefont {A.}~\bibnamefont {Zeilinger}},\ }\href {\doibase
  10.1038/nphoton.2016.12} {\bibfield  {journal} {\bibinfo  {journal} {Nature
  Photonics}\ }\textbf {\bibinfo {volume} {10}},\ \bibinfo {pages} {248}
  (\bibinfo {year} {2016})}\BibitemShut {NoStop}%
\bibitem [{\citenamefont {Sit}\ \emph {et~al.}(2017)\citenamefont {Sit},
  \citenamefont {Bouchard}, \citenamefont {Fickler}, \citenamefont
  {Gagnon-Bischoff}, \citenamefont {Larocque}, \citenamefont {Heshami},
  \citenamefont {Elser}, \citenamefont {Peuntinger}, \citenamefont
  {G\"{u}nthner}, \citenamefont {Heim}, \citenamefont {Marquardt},
  \citenamefont {Leuchs}, \citenamefont {Boyd},\ and\ \citenamefont
  {Karimi}}]{Sit17}%
  \BibitemOpen
  \bibfield  {author} {\bibinfo {author} {\bibfnamefont {A.}~\bibnamefont
  {Sit}}, \bibinfo {author} {\bibfnamefont {F.}~\bibnamefont {Bouchard}},
  \bibinfo {author} {\bibfnamefont {R.}~\bibnamefont {Fickler}}, \bibinfo
  {author} {\bibfnamefont {J.}~\bibnamefont {Gagnon-Bischoff}}, \bibinfo
  {author} {\bibfnamefont {H.}~\bibnamefont {Larocque}}, \bibinfo {author}
  {\bibfnamefont {K.}~\bibnamefont {Heshami}}, \bibinfo {author} {\bibfnamefont
  {D.}~\bibnamefont {Elser}}, \bibinfo {author} {\bibfnamefont
  {C.}~\bibnamefont {Peuntinger}}, \bibinfo {author} {\bibfnamefont
  {K.}~\bibnamefont {G\"{u}nthner}}, \bibinfo {author} {\bibfnamefont
  {B.}~\bibnamefont {Heim}}, \bibinfo {author} {\bibfnamefont {C.}~\bibnamefont
  {Marquardt}}, \bibinfo {author} {\bibfnamefont {G.}~\bibnamefont {Leuchs}},
  \bibinfo {author} {\bibfnamefont {R.~W.}\ \bibnamefont {Boyd}}, \ and\
  \bibinfo {author} {\bibfnamefont {E.}~\bibnamefont {Karimi}},\ }\href
  {\doibase 10.1364/OPTICA.4.001006} {\bibfield  {journal} {\bibinfo  {journal}
  {Optica}\ }\textbf {\bibinfo {volume} {4}},\ \bibinfo {pages} {1006}
  (\bibinfo {year} {2017})}\BibitemShut {NoStop}%
\bibitem [{\citenamefont {Cozzolino}\ \emph
  {et~al.}(2019{\natexlab{c}})\citenamefont {Cozzolino}, \citenamefont {Bacco},
  \citenamefont {Da~Lio}, \citenamefont {Ingerslev}, \citenamefont {Ding},
  \citenamefont {Dalgaard}, \citenamefont {Kristensen}, \citenamefont {Galili},
  \citenamefont {Rottwitt}, \citenamefont {Ramachandran},\ and\ \citenamefont
  {Oxenl\o{}we}}]{Cozzolino2019_fiber}%
  \BibitemOpen
  \bibfield  {author} {\bibinfo {author} {\bibfnamefont {D.}~\bibnamefont
  {Cozzolino}}, \bibinfo {author} {\bibfnamefont {D.}~\bibnamefont {Bacco}},
  \bibinfo {author} {\bibfnamefont {B.}~\bibnamefont {Da~Lio}}, \bibinfo
  {author} {\bibfnamefont {K.}~\bibnamefont {Ingerslev}}, \bibinfo {author}
  {\bibfnamefont {Y.}~\bibnamefont {Ding}}, \bibinfo {author} {\bibfnamefont
  {K.}~\bibnamefont {Dalgaard}}, \bibinfo {author} {\bibfnamefont
  {P.}~\bibnamefont {Kristensen}}, \bibinfo {author} {\bibfnamefont
  {M.}~\bibnamefont {Galili}}, \bibinfo {author} {\bibfnamefont
  {K.}~\bibnamefont {Rottwitt}}, \bibinfo {author} {\bibfnamefont
  {S.}~\bibnamefont {Ramachandran}}, \ and\ \bibinfo {author} {\bibfnamefont
  {L.~K.}\ \bibnamefont {Oxenl\o{}we}},\ }\href {\doibase
  10.1103/PhysRevApplied.11.064058} {\bibfield  {journal} {\bibinfo  {journal}
  {Phys. Rev. Appl.}\ }\textbf {\bibinfo {volume} {11}},\ \bibinfo {pages}
  {064058} (\bibinfo {year} {2019}{\natexlab{c}})}\BibitemShut {NoStop}%
\bibitem [{\citenamefont {Zhang}\ \emph {et~al.}(2010)\citenamefont {Zhang},
  \citenamefont {Liu}, \citenamefont {Liu}, \citenamefont {Li}, \citenamefont
  {Li},\ and\ \citenamefont {Guo}}]{zhang-oam-qw-2010}%
  \BibitemOpen
  \bibfield  {author} {\bibinfo {author} {\bibfnamefont {P.}~\bibnamefont
  {Zhang}}, \bibinfo {author} {\bibfnamefont {B.~H.}\ \bibnamefont {Liu}},
  \bibinfo {author} {\bibfnamefont {R.~F.}\ \bibnamefont {Liu}}, \bibinfo
  {author} {\bibfnamefont {H.~R.}\ \bibnamefont {Li}}, \bibinfo {author}
  {\bibfnamefont {F.~L.}\ \bibnamefont {Li}}, \ and\ \bibinfo {author}
  {\bibfnamefont {G.~C.}\ \bibnamefont {Guo}},\ }\href
  {https://journals.aps.org/pra/abstract/10.1103/PhysRevA.81.052322} {\bibfield
   {journal} {\bibinfo  {journal} {Phys. Rev. A}\ }\textbf {\bibinfo {volume}
  {81}},\ \bibinfo {pages} {052322} (\bibinfo {year} {2010})}\BibitemShut
  {NoStop}%
\bibitem [{\citenamefont {Goyal}\ \emph {et~al.}(2013)\citenamefont {Goyal},
  \citenamefont {Roux}, \citenamefont {Forbes},\ and\ \citenamefont
  {Konrad}}]{goyal2013implementing}%
  \BibitemOpen
  \bibfield  {author} {\bibinfo {author} {\bibfnamefont {S.~K.}\ \bibnamefont
  {Goyal}}, \bibinfo {author} {\bibfnamefont {F.~S.}\ \bibnamefont {Roux}},
  \bibinfo {author} {\bibfnamefont {A.}~\bibnamefont {Forbes}}, \ and\ \bibinfo
  {author} {\bibfnamefont {T.}~\bibnamefont {Konrad}},\ }\href {\doibase
  10.1103/physrevlett.110.263602} {\bibfield  {journal} {\bibinfo  {journal}
  {Phys. Rev. Lett.}\ }\textbf {\bibinfo {volume} {110}},\ \bibinfo {pages}
  {263602} (\bibinfo {year} {2013})}\BibitemShut {NoStop}%
\bibitem [{\citenamefont {Cardano}\ \emph {et~al.}(2015)\citenamefont
  {Cardano}, \citenamefont {Massa}, \citenamefont {Qassim}, \citenamefont
  {Karimi}, \citenamefont {Slussarenko}, \citenamefont {Paparo}, \citenamefont
  {de~Lisio}, \citenamefont {Sciarrino}, \citenamefont {Santamato},
  \citenamefont {Boyd},\ and\ \citenamefont {Marrucci}}]{cardano2015quantum}%
  \BibitemOpen
  \bibfield  {author} {\bibinfo {author} {\bibfnamefont {F.}~\bibnamefont
  {Cardano}}, \bibinfo {author} {\bibfnamefont {F.}~\bibnamefont {Massa}},
  \bibinfo {author} {\bibfnamefont {H.}~\bibnamefont {Qassim}}, \bibinfo
  {author} {\bibfnamefont {E.}~\bibnamefont {Karimi}}, \bibinfo {author}
  {\bibfnamefont {S.}~\bibnamefont {Slussarenko}}, \bibinfo {author}
  {\bibfnamefont {D.}~\bibnamefont {Paparo}}, \bibinfo {author} {\bibfnamefont
  {C.}~\bibnamefont {de~Lisio}}, \bibinfo {author} {\bibfnamefont
  {F.}~\bibnamefont {Sciarrino}}, \bibinfo {author} {\bibfnamefont
  {E.}~\bibnamefont {Santamato}}, \bibinfo {author} {\bibfnamefont {R.~W.}\
  \bibnamefont {Boyd}}, \ and\ \bibinfo {author} {\bibfnamefont
  {L.}~\bibnamefont {Marrucci}},\ }\href {\doibase 10.1126/sciadv.1500087}
  {\bibfield  {journal} {\bibinfo  {journal} {Science Advances}\ }\textbf
  {\bibinfo {volume} {1}},\ \bibinfo {pages} {e1500087} (\bibinfo {year}
  {2015})}\BibitemShut {NoStop}%
\bibitem [{\citenamefont {Cardano}\ \emph {et~al.}(2016)\citenamefont
  {Cardano}, \citenamefont {Maffei}, \citenamefont {Massa}, \citenamefont
  {Piccirillo}, \citenamefont {de~Lisio}, \citenamefont {Filippis},
  \citenamefont {Cataudella}, \citenamefont {Santamato},\ and\ \citenamefont
  {Marrucci}}]{cardano2016statistical}%
  \BibitemOpen
  \bibfield  {author} {\bibinfo {author} {\bibfnamefont {F.}~\bibnamefont
  {Cardano}}, \bibinfo {author} {\bibfnamefont {M.}~\bibnamefont {Maffei}},
  \bibinfo {author} {\bibfnamefont {F.}~\bibnamefont {Massa}}, \bibinfo
  {author} {\bibfnamefont {B.}~\bibnamefont {Piccirillo}}, \bibinfo {author}
  {\bibfnamefont {C.}~\bibnamefont {de~Lisio}}, \bibinfo {author}
  {\bibfnamefont {G.~D.}\ \bibnamefont {Filippis}}, \bibinfo {author}
  {\bibfnamefont {V.}~\bibnamefont {Cataudella}}, \bibinfo {author}
  {\bibfnamefont {E.}~\bibnamefont {Santamato}}, \ and\ \bibinfo {author}
  {\bibfnamefont {L.}~\bibnamefont {Marrucci}},\ }\href {\doibase
  10.1038/ncomms11439} {\bibfield  {journal} {\bibinfo  {journal} {Nat. Comm.}\
  }\textbf {\bibinfo {volume} {7}},\ \bibinfo {pages} {11439} (\bibinfo {year}
  {2016})}\BibitemShut {NoStop}%
\bibitem [{\citenamefont {Cardano}\ \emph {et~al.}(2017)\citenamefont
  {Cardano}, \citenamefont {D'Errico}, \citenamefont {Dauphin}, \citenamefont
  {Maffei}, \citenamefont {Piccirillo}, \citenamefont {de~Lisio}, \citenamefont
  {Filippis}, \citenamefont {Cataudella}, \citenamefont {Santamato},
  \citenamefont {Marrucci}, \citenamefont {Lewenstein},\ and\ \citenamefont
  {Massignan}}]{cardano_zak_2017}%
  \BibitemOpen
  \bibfield  {author} {\bibinfo {author} {\bibfnamefont {F.}~\bibnamefont
  {Cardano}}, \bibinfo {author} {\bibfnamefont {A.}~\bibnamefont {D'Errico}},
  \bibinfo {author} {\bibfnamefont {A.}~\bibnamefont {Dauphin}}, \bibinfo
  {author} {\bibfnamefont {M.}~\bibnamefont {Maffei}}, \bibinfo {author}
  {\bibfnamefont {B.}~\bibnamefont {Piccirillo}}, \bibinfo {author}
  {\bibfnamefont {C.}~\bibnamefont {de~Lisio}}, \bibinfo {author}
  {\bibfnamefont {G.~D.}\ \bibnamefont {Filippis}}, \bibinfo {author}
  {\bibfnamefont {V.}~\bibnamefont {Cataudella}}, \bibinfo {author}
  {\bibfnamefont {E.}~\bibnamefont {Santamato}}, \bibinfo {author}
  {\bibfnamefont {L.}~\bibnamefont {Marrucci}}, \bibinfo {author}
  {\bibfnamefont {M.}~\bibnamefont {Lewenstein}}, \ and\ \bibinfo {author}
  {\bibfnamefont {P.}~\bibnamefont {Massignan}},\ }\href
  {https://www.nature.com/articles/ncomms15516} {\bibfield  {journal} {\bibinfo
   {journal} {Nat. Comm.}\ }\textbf {\bibinfo {volume} {8}},\ \bibinfo {pages}
  {15516} (\bibinfo {year} {2017})}\BibitemShut {NoStop}%
\bibitem [{\citenamefont {Innocenti}\ \emph {et~al.}(2017)\citenamefont
  {Innocenti}, \citenamefont {Majury}, \citenamefont {Giordani}, \citenamefont
  {Spagnolo}, \citenamefont {Sciarrino}, \citenamefont {Paternostro},\ and\
  \citenamefont {Ferraro}}]{Innocenti2017}%
  \BibitemOpen
  \bibfield  {author} {\bibinfo {author} {\bibfnamefont {L.}~\bibnamefont
  {Innocenti}}, \bibinfo {author} {\bibfnamefont {H.}~\bibnamefont {Majury}},
  \bibinfo {author} {\bibfnamefont {T.}~\bibnamefont {Giordani}}, \bibinfo
  {author} {\bibfnamefont {N.}~\bibnamefont {Spagnolo}}, \bibinfo {author}
  {\bibfnamefont {F.}~\bibnamefont {Sciarrino}}, \bibinfo {author}
  {\bibfnamefont {M.}~\bibnamefont {Paternostro}}, \ and\ \bibinfo {author}
  {\bibfnamefont {A.}~\bibnamefont {Ferraro}},\ }\href {\doibase
  10.1103/PhysRevA.96.062326} {\bibfield  {journal} {\bibinfo  {journal} {Phys.
  Rev. A}\ }\textbf {\bibinfo {volume} {96}},\ \bibinfo {pages} {062326}
  (\bibinfo {year} {2017})}\BibitemShut {NoStop}%
\bibitem [{\citenamefont {Giordani}\ \emph {et~al.}(2019)\citenamefont
  {Giordani}, \citenamefont {Polino}, \citenamefont {Emiliani}, \citenamefont
  {Suprano}, \citenamefont {Innocenti}, \citenamefont {Majury}, \citenamefont
  {Marrucci}, \citenamefont {Paternostro}, \citenamefont {Ferraro},
  \citenamefont {Spagnolo},\ and\ \citenamefont {Sciarrino}}]{giordani_2018}%
  \BibitemOpen
  \bibfield  {author} {\bibinfo {author} {\bibfnamefont {T.}~\bibnamefont
  {Giordani}}, \bibinfo {author} {\bibfnamefont {E.}~\bibnamefont {Polino}},
  \bibinfo {author} {\bibfnamefont {S.}~\bibnamefont {Emiliani}}, \bibinfo
  {author} {\bibfnamefont {A.}~\bibnamefont {Suprano}}, \bibinfo {author}
  {\bibfnamefont {L.}~\bibnamefont {Innocenti}}, \bibinfo {author}
  {\bibfnamefont {H.}~\bibnamefont {Majury}}, \bibinfo {author} {\bibfnamefont
  {L.}~\bibnamefont {Marrucci}}, \bibinfo {author} {\bibfnamefont
  {M.}~\bibnamefont {Paternostro}}, \bibinfo {author} {\bibfnamefont
  {A.}~\bibnamefont {Ferraro}}, \bibinfo {author} {\bibfnamefont
  {N.}~\bibnamefont {Spagnolo}}, \ and\ \bibinfo {author} {\bibfnamefont
  {F.}~\bibnamefont {Sciarrino}},\ }\href {\doibase
  10.1103/PhysRevLett.122.020503} {\bibfield  {journal} {\bibinfo  {journal}
  {Phys. Rev. Lett.}\ }\textbf {\bibinfo {volume} {122}},\ \bibinfo {pages}
  {020503} (\bibinfo {year} {2019})}\BibitemShut {NoStop}%
\bibitem [{\citenamefont {Leach}\ \emph {et~al.}(2002)\citenamefont {Leach},
  \citenamefont {Padgett}, \citenamefont {Barnett}, \citenamefont
  {Franke-Arnold},\ and\ \citenamefont {Courtial}}]{Leach2002_oamSorter}%
  \BibitemOpen
  \bibfield  {author} {\bibinfo {author} {\bibfnamefont {J.}~\bibnamefont
  {Leach}}, \bibinfo {author} {\bibfnamefont {M.~J.}\ \bibnamefont {Padgett}},
  \bibinfo {author} {\bibfnamefont {S.~M.}\ \bibnamefont {Barnett}}, \bibinfo
  {author} {\bibfnamefont {S.}~\bibnamefont {Franke-Arnold}}, \ and\ \bibinfo
  {author} {\bibfnamefont {J.}~\bibnamefont {Courtial}},\ }\href {\doibase
  10.1103/PhysRevLett.88.257901} {\bibfield  {journal} {\bibinfo  {journal}
  {Phys. Rev. Lett.}\ }\textbf {\bibinfo {volume} {88}},\ \bibinfo {pages}
  {257901} (\bibinfo {year} {2002})}\BibitemShut {NoStop}%
\bibitem [{\citenamefont {Slussarenko}\ \emph {et~al.}(2010)\citenamefont
  {Slussarenko}, \citenamefont {D'Ambrosio}, \citenamefont {Piccirillo},
  \citenamefont {Marrucci},\ and\ \citenamefont
  {Santamato}}]{Slussarenko2010_oamSorter}%
  \BibitemOpen
  \bibfield  {author} {\bibinfo {author} {\bibfnamefont {S.}~\bibnamefont
  {Slussarenko}}, \bibinfo {author} {\bibfnamefont {V.}~\bibnamefont
  {D'Ambrosio}}, \bibinfo {author} {\bibfnamefont {B.}~\bibnamefont
  {Piccirillo}}, \bibinfo {author} {\bibfnamefont {L.}~\bibnamefont
  {Marrucci}}, \ and\ \bibinfo {author} {\bibfnamefont {E.}~\bibnamefont
  {Santamato}},\ }\href {\doibase 10.1364/OE.18.027205} {\bibfield  {journal}
  {\bibinfo  {journal} {Opt. Express}\ }\textbf {\bibinfo {volume} {18}},\
  \bibinfo {pages} {27205} (\bibinfo {year} {2010})}\BibitemShut {NoStop}%
\bibitem [{\citenamefont {Bauer}\ \emph {et~al.}(2014)\citenamefont {Bauer},
  \citenamefont {Orlov}, \citenamefont {Peschel}, \citenamefont {Banzer},\ and\
  \citenamefont {Leuchs}}]{Bauer2014}%
  \BibitemOpen
  \bibfield  {author} {\bibinfo {author} {\bibfnamefont {T.}~\bibnamefont
  {Bauer}}, \bibinfo {author} {\bibfnamefont {S.}~\bibnamefont {Orlov}},
  \bibinfo {author} {\bibfnamefont {U.}~\bibnamefont {Peschel}}, \bibinfo
  {author} {\bibfnamefont {P.}~\bibnamefont {Banzer}}, \ and\ \bibinfo {author}
  {\bibfnamefont {G.}~\bibnamefont {Leuchs}},\ }\href {\doibase
  10.1038/NPHOTON.2013.289} {\bibfield  {journal} {\bibinfo  {journal} {Nature
  Photonics}\ }\textbf {\bibinfo {volume} {8}},\ \bibinfo {pages} {23}
  (\bibinfo {year} {2014})}\BibitemShut {NoStop}%
\bibitem [{\citenamefont {Berkhout}\ \emph {et~al.}(2010)\citenamefont
  {Berkhout}, \citenamefont {Lavery}, \citenamefont {Courtial}, \citenamefont
  {Beijersbergen},\ and\ \citenamefont {Padgett}}]{Berkhout2010_oamSorter}%
  \BibitemOpen
  \bibfield  {author} {\bibinfo {author} {\bibfnamefont {G.~C.~G.}\
  \bibnamefont {Berkhout}}, \bibinfo {author} {\bibfnamefont {M.~P.~J.}\
  \bibnamefont {Lavery}}, \bibinfo {author} {\bibfnamefont {J.}~\bibnamefont
  {Courtial}}, \bibinfo {author} {\bibfnamefont {M.~W.}\ \bibnamefont
  {Beijersbergen}}, \ and\ \bibinfo {author} {\bibfnamefont {M.~J.}\
  \bibnamefont {Padgett}},\ }\href {\doibase 10.1103/PhysRevLett.105.153601}
  {\bibfield  {journal} {\bibinfo  {journal} {Phys. Rev. Lett.}\ }\textbf
  {\bibinfo {volume} {105}},\ \bibinfo {pages} {153601} (\bibinfo {year}
  {2010})}\BibitemShut {NoStop}%
\bibitem [{\citenamefont {Bolduc}\ \emph {et~al.}(2013)\citenamefont {Bolduc},
  \citenamefont {Bent}, \citenamefont {Santamato}, \citenamefont {Karimi},\
  and\ \citenamefont {Boyd}}]{bolduc2013holo}%
  \BibitemOpen
  \bibfield  {author} {\bibinfo {author} {\bibfnamefont {E.}~\bibnamefont
  {Bolduc}}, \bibinfo {author} {\bibfnamefont {N.}~\bibnamefont {Bent}},
  \bibinfo {author} {\bibfnamefont {E.}~\bibnamefont {Santamato}}, \bibinfo
  {author} {\bibfnamefont {E.}~\bibnamefont {Karimi}}, \ and\ \bibinfo {author}
  {\bibfnamefont {R.~W.}\ \bibnamefont {Boyd}},\ }\href {\doibase
  10.1364/OL.38.003546} {\bibfield  {journal} {\bibinfo  {journal} {Optics
  Letters}\ }\textbf {\bibinfo {volume} {38}},\ \bibinfo {pages} {3546}
  (\bibinfo {year} {2013})}\BibitemShut {NoStop}%
\bibitem [{\citenamefont {Malik}\ \emph {et~al.}(2014)\citenamefont {Malik},
  \citenamefont {Mirhosseini}, \citenamefont {Lavery}, \citenamefont {Leach},\
  and\ \citenamefont {Boyd}}]{Mehul2014_oamSorter}%
  \BibitemOpen
  \bibfield  {author} {\bibinfo {author} {\bibfnamefont {M.}~\bibnamefont
  {Malik}}, \bibinfo {author} {\bibfnamefont {M.}~\bibnamefont {Mirhosseini}},
  \bibinfo {author} {\bibfnamefont {M.~P.~J.}\ \bibnamefont {Lavery}}, \bibinfo
  {author} {\bibfnamefont {M.~J.}\ \bibnamefont {Leach}, \bibfnamefont
  {Jonathan abd~Padgett}}, \ and\ \bibinfo {author} {\bibfnamefont {R.~W.}\
  \bibnamefont {Boyd}},\ }\href {https://doi.org/10.1038/ncomms4115} {\bibfield
   {journal} {\bibinfo  {journal} {Nature Communications}\ }\textbf {\bibinfo
  {volume} {5}} (\bibinfo {year} {2014})}\BibitemShut {NoStop}%
\bibitem [{\citenamefont {Qassim}\ \emph {et~al.}(2014)\citenamefont {Qassim},
  \citenamefont {Miatto}, \citenamefont {Torres}, \citenamefont {Padgett},
  \citenamefont {Karimi},\ and\ \citenamefont {Boyd}}]{Qassim2014}%
  \BibitemOpen
  \bibfield  {author} {\bibinfo {author} {\bibfnamefont {H.}~\bibnamefont
  {Qassim}}, \bibinfo {author} {\bibfnamefont {F.~M.}\ \bibnamefont {Miatto}},
  \bibinfo {author} {\bibfnamefont {J.~P.}\ \bibnamefont {Torres}}, \bibinfo
  {author} {\bibfnamefont {M.~J.}\ \bibnamefont {Padgett}}, \bibinfo {author}
  {\bibfnamefont {E.}~\bibnamefont {Karimi}}, \ and\ \bibinfo {author}
  {\bibfnamefont {R.~W.}\ \bibnamefont {Boyd}},\ }\href {\doibase
  10.1364/JOSAB.31.000A20} {\bibfield  {journal} {\bibinfo  {journal} {J. Opt.
  Soc. Am. B}\ }\textbf {\bibinfo {volume} {31}},\ \bibinfo {pages} {A20}
  (\bibinfo {year} {2014})}\BibitemShut {NoStop}%
\bibitem [{\citenamefont {Paris}\ and\ \citenamefont
  {\v{R}eh\'{a}\v{c}ek}(2004)}]{Paris2004}%
  \BibitemOpen
  \bibfield  {author} {\bibinfo {author} {\bibfnamefont {M.~G.~A.}\
  \bibnamefont {Paris}}\ and\ \bibinfo {author} {\bibfnamefont
  {J.}~\bibnamefont {\v{R}eh\'{a}\v{c}ek}},\ }\href {\doibase 10.1007/b98673}
  {\emph {\bibinfo {title} {Quantum State Estimation}}}\ (\bibinfo  {publisher}
  {Lecture Notes in Physics, vol. 649, SPRINGER},\ \bibinfo {year}
  {2004})\BibitemShut {NoStop}%
\bibitem [{\citenamefont {Banaszek}\ \emph {et~al.}(2013)\citenamefont
  {Banaszek}, \citenamefont {Cramer},\ and\ \citenamefont
  {Gross}}]{Banaszek_2013}%
  \BibitemOpen
  \bibfield  {author} {\bibinfo {author} {\bibfnamefont {K.}~\bibnamefont
  {Banaszek}}, \bibinfo {author} {\bibfnamefont {M.}~\bibnamefont {Cramer}}, \
  and\ \bibinfo {author} {\bibfnamefont {D.}~\bibnamefont {Gross}},\ }\href
  {\doibase 10.1088/1367-2630/15/12/125020} {\bibfield  {journal} {\bibinfo
  {journal} {New Journal of Physics}\ }\textbf {\bibinfo {volume} {15}},\
  \bibinfo {pages} {125020} (\bibinfo {year} {2013})}\BibitemShut {NoStop}%
\bibitem [{\citenamefont {Zhenxing}\ \emph {et~al.}(2017)\citenamefont
  {Zhenxing}, \citenamefont {Yuanyuan}, \citenamefont {Yougang}, \citenamefont
  {Yachao}, \citenamefont {Weixing}, \citenamefont {Hailu},\ and\ \citenamefont
  {Shuangchun}}]{Liu:17}%
  \BibitemOpen
  \bibfield  {author} {\bibinfo {author} {\bibfnamefont {L.}~\bibnamefont
  {Zhenxing}}, \bibinfo {author} {\bibfnamefont {L.}~\bibnamefont {Yuanyuan}},
  \bibinfo {author} {\bibfnamefont {K.}~\bibnamefont {Yougang}}, \bibinfo
  {author} {\bibfnamefont {L.}~\bibnamefont {Yachao}}, \bibinfo {author}
  {\bibfnamefont {S.}~\bibnamefont {Weixing}}, \bibinfo {author} {\bibfnamefont
  {L.}~\bibnamefont {Hailu}}, \ and\ \bibinfo {author} {\bibfnamefont
  {W.}~\bibnamefont {Shuangchun}},\ }\href {\doibase 10.1364/PRJ.5.000015}
  {\bibfield  {journal} {\bibinfo  {journal} {Photon. Res.}\ }\textbf {\bibinfo
  {volume} {5}},\ \bibinfo {pages} {15} (\bibinfo {year} {2017})}\BibitemShut
  {NoStop}%
\bibitem [{\citenamefont {Ndagano}\ \emph {et~al.}(2018)\citenamefont
  {Ndagano}, \citenamefont {Nape}, \citenamefont {Cox}, \citenamefont
  {Rosales-Guzman},\ and\ \citenamefont {Forbes}}]{Ndagano:18}%
  \BibitemOpen
  \bibfield  {author} {\bibinfo {author} {\bibfnamefont {B.}~\bibnamefont
  {Ndagano}}, \bibinfo {author} {\bibfnamefont {I.}~\bibnamefont {Nape}},
  \bibinfo {author} {\bibfnamefont {M.~A.}\ \bibnamefont {Cox}}, \bibinfo
  {author} {\bibfnamefont {C.}~\bibnamefont {Rosales-Guzman}}, \ and\ \bibinfo
  {author} {\bibfnamefont {A.}~\bibnamefont {Forbes}},\ }\href
  {http://jlt.osa.org/abstract.cfm?URI=jlt-36-2-292} {\bibfield  {journal}
  {\bibinfo  {journal} {J. Lightwave Technol.}\ }\textbf {\bibinfo {volume}
  {36}},\ \bibinfo {pages} {292} (\bibinfo {year} {2018})}\BibitemShut
  {NoStop}%
\bibitem [{\citenamefont {Chen}\ \emph {et~al.}(2018)\citenamefont {Chen},
  \citenamefont {Gao}, \citenamefont {Jiao}, \citenamefont {Sun}, \citenamefont
  {Shen}, \citenamefont {Qiao}, \citenamefont {Tang}, \citenamefont {Lin},\
  and\ \citenamefont {Jin}}]{oamchip}%
  \BibitemOpen
  \bibfield  {author} {\bibinfo {author} {\bibfnamefont {Y.}~\bibnamefont
  {Chen}}, \bibinfo {author} {\bibfnamefont {J.}~\bibnamefont {Gao}}, \bibinfo
  {author} {\bibfnamefont {Z.-Q.}\ \bibnamefont {Jiao}}, \bibinfo {author}
  {\bibfnamefont {K.}~\bibnamefont {Sun}}, \bibinfo {author} {\bibfnamefont
  {W.-G.}\ \bibnamefont {Shen}}, \bibinfo {author} {\bibfnamefont {L.-F.}\
  \bibnamefont {Qiao}}, \bibinfo {author} {\bibfnamefont {H.}~\bibnamefont
  {Tang}}, \bibinfo {author} {\bibfnamefont {X.-F.}\ \bibnamefont {Lin}}, \
  and\ \bibinfo {author} {\bibfnamefont {X.-M.}\ \bibnamefont {Jin}},\ }\href
  {\doibase 10.1103/PhysRevLett.121.233602} {\bibfield  {journal} {\bibinfo
  {journal} {Physical Review Letters}\ }\textbf {\bibinfo {volume} {121}},\
  \bibinfo {pages} {233602} (\bibinfo {year} {2018})}\BibitemShut {NoStop}%
\bibitem [{\citenamefont {Cai}\ \emph {et~al.}(2012)\citenamefont {Cai},
  \citenamefont {Wang}, \citenamefont {Strain}, \citenamefont {Johnson-Morris},
  \citenamefont {Zhu}, \citenamefont {Sorel}, \citenamefont {O'Brien},
  \citenamefont {Thompson},\ and\ \citenamefont {Yu}}]{cai2012integrated}%
  \BibitemOpen
  \bibfield  {author} {\bibinfo {author} {\bibfnamefont {X.}~\bibnamefont
  {Cai}}, \bibinfo {author} {\bibfnamefont {J.}~\bibnamefont {Wang}}, \bibinfo
  {author} {\bibfnamefont {M.~J.}\ \bibnamefont {Strain}}, \bibinfo {author}
  {\bibfnamefont {B.}~\bibnamefont {Johnson-Morris}}, \bibinfo {author}
  {\bibfnamefont {J.}~\bibnamefont {Zhu}}, \bibinfo {author} {\bibfnamefont
  {M.}~\bibnamefont {Sorel}}, \bibinfo {author} {\bibfnamefont {J.~L.}\
  \bibnamefont {O'Brien}}, \bibinfo {author} {\bibfnamefont {M.~G.}\
  \bibnamefont {Thompson}}, \ and\ \bibinfo {author} {\bibfnamefont
  {S.}~\bibnamefont {Yu}},\ }\href {\doibase 10.1126/science.1226528}
  {\bibfield  {journal} {\bibinfo  {journal} {Science}\ }\textbf {\bibinfo
  {volume} {338}},\ \bibinfo {pages} {363} (\bibinfo {year}
  {2012})}\BibitemShut {NoStop}%
\bibitem [{\citenamefont {Liu}\ \emph {et~al.}(2018)\citenamefont {Liu},
  \citenamefont {Li}, \citenamefont {Zhu}, \citenamefont {Wang}, \citenamefont
  {Chen}, \citenamefont {Klitis}, \citenamefont {Du}, \citenamefont {Mo},
  \citenamefont {Sorel}, \citenamefont {Yu}, \citenamefont {Cai},\ and\
  \citenamefont {Wang}}]{chiptofiber}%
  \BibitemOpen
  \bibfield  {author} {\bibinfo {author} {\bibfnamefont {J.}~\bibnamefont
  {Liu}}, \bibinfo {author} {\bibfnamefont {S.-M.}\ \bibnamefont {Li}},
  \bibinfo {author} {\bibfnamefont {L.}~\bibnamefont {Zhu}}, \bibinfo {author}
  {\bibfnamefont {A.-D.}\ \bibnamefont {Wang}}, \bibinfo {author}
  {\bibfnamefont {S.}~\bibnamefont {Chen}}, \bibinfo {author} {\bibfnamefont
  {C.}~\bibnamefont {Klitis}}, \bibinfo {author} {\bibfnamefont
  {C.}~\bibnamefont {Du}}, \bibinfo {author} {\bibfnamefont {Q.}~\bibnamefont
  {Mo}}, \bibinfo {author} {\bibfnamefont {M.}~\bibnamefont {Sorel}}, \bibinfo
  {author} {\bibfnamefont {S.-Y.}\ \bibnamefont {Yu}}, \bibinfo {author}
  {\bibfnamefont {X.-L.}\ \bibnamefont {Cai}}, \ and\ \bibinfo {author}
  {\bibfnamefont {J.}~\bibnamefont {Wang}},\ }\href {\doibase
  10.1038/lsa.2017.148} {\bibfield  {journal} {\bibinfo  {journal} {Light:
  Science \& Applications}\ }\textbf {\bibinfo {volume} {7}},\ \bibinfo {pages}
  {17148} (\bibinfo {year} {2018})}\BibitemShut {NoStop}%
\bibitem [{\citenamefont {Karimi}\ \emph {et~al.}(2014)\citenamefont {Karimi},
  \citenamefont {Schulz}, \citenamefont {De~Leon}, \citenamefont {Qassim},
  \citenamefont {Upham},\ and\ \citenamefont {Boyd}}]{karimi2014}%
  \BibitemOpen
  \bibfield  {author} {\bibinfo {author} {\bibfnamefont {E.}~\bibnamefont
  {Karimi}}, \bibinfo {author} {\bibfnamefont {S.~A.}\ \bibnamefont {Schulz}},
  \bibinfo {author} {\bibfnamefont {I.}~\bibnamefont {De~Leon}}, \bibinfo
  {author} {\bibfnamefont {H.}~\bibnamefont {Qassim}}, \bibinfo {author}
  {\bibfnamefont {J.}~\bibnamefont {Upham}}, \ and\ \bibinfo {author}
  {\bibfnamefont {R.~W.}\ \bibnamefont {Boyd}},\ }\href
  {https://doi.org/10.1038/lsa.2014.48} {\bibfield  {journal} {\bibinfo
  {journal} {Light: Science \& Applications}\ }\textbf {\bibinfo {volume}
  {3}},\ \bibinfo {pages} {e167} (\bibinfo {year} {2014})}\BibitemShut
  {NoStop}%
\bibitem [{\citenamefont {Yue}\ \emph {et~al.}(2016)\citenamefont {Yue},
  \citenamefont {Wen}, \citenamefont {Xin}, \citenamefont {Gerardot},
  \citenamefont {Li},\ and\ \citenamefont {Chen}}]{yue2016}%
  \BibitemOpen
  \bibfield  {author} {\bibinfo {author} {\bibfnamefont {F.}~\bibnamefont
  {Yue}}, \bibinfo {author} {\bibfnamefont {D.}~\bibnamefont {Wen}}, \bibinfo
  {author} {\bibfnamefont {J.}~\bibnamefont {Xin}}, \bibinfo {author}
  {\bibfnamefont {B.~D.}\ \bibnamefont {Gerardot}}, \bibinfo {author}
  {\bibfnamefont {J.}~\bibnamefont {Li}}, \ and\ \bibinfo {author}
  {\bibfnamefont {X.}~\bibnamefont {Chen}},\ }\href
  {https://doi.org/10.1021/acsphotonics.6b00392} {\bibfield  {journal}
  {\bibinfo  {journal} {ACS photonics}\ }\textbf {\bibinfo {volume} {3}},\
  \bibinfo {pages} {1558} (\bibinfo {year} {2016})}\BibitemShut {NoStop}%
\bibitem [{\citenamefont {Carrasquilla}\ \emph {et~al.}(2019)\citenamefont
  {Carrasquilla}, \citenamefont {Torlai}, \citenamefont {Melko},\ and\
  \citenamefont {Aolita}}]{carrasquilla2019reconstructing}%
  \BibitemOpen
  \bibfield  {author} {\bibinfo {author} {\bibfnamefont {J.}~\bibnamefont
  {Carrasquilla}}, \bibinfo {author} {\bibfnamefont {G.}~\bibnamefont
  {Torlai}}, \bibinfo {author} {\bibfnamefont {R.~G.}\ \bibnamefont {Melko}}, \
  and\ \bibinfo {author} {\bibfnamefont {L.}~\bibnamefont {Aolita}},\ }\href
  {\doibase 10.1038/s42256-019-0028-1} {\bibfield  {journal} {\bibinfo
  {journal} {Nat. Mach. Intell.}\ }\textbf {\bibinfo {volume} {1}},\ \bibinfo
  {pages} {155} (\bibinfo {year} {2019})}\BibitemShut {NoStop}%
\bibitem [{\citenamefont {Giordani}\ \emph {et~al.}(2018)\citenamefont
  {Giordani}, \citenamefont {Flamini}, \citenamefont {Pompili}, \citenamefont
  {Viggianiello}, \citenamefont {Spagnolo}, \citenamefont {Crespi},
  \citenamefont {Osellame}, \citenamefont {Wiebe}, \citenamefont {Walschaers},
  \citenamefont {Buchleitner} \emph {et~al.}}]{tairelly}%
  \BibitemOpen
  \bibfield  {author} {\bibinfo {author} {\bibfnamefont {T.}~\bibnamefont
  {Giordani}}, \bibinfo {author} {\bibfnamefont {F.}~\bibnamefont {Flamini}},
  \bibinfo {author} {\bibfnamefont {M.}~\bibnamefont {Pompili}}, \bibinfo
  {author} {\bibfnamefont {N.}~\bibnamefont {Viggianiello}}, \bibinfo {author}
  {\bibfnamefont {N.}~\bibnamefont {Spagnolo}}, \bibinfo {author}
  {\bibfnamefont {A.}~\bibnamefont {Crespi}}, \bibinfo {author} {\bibfnamefont
  {R.}~\bibnamefont {Osellame}}, \bibinfo {author} {\bibfnamefont
  {N.}~\bibnamefont {Wiebe}}, \bibinfo {author} {\bibfnamefont
  {M.}~\bibnamefont {Walschaers}}, \bibinfo {author} {\bibfnamefont
  {A.}~\bibnamefont {Buchleitner}},  \emph {et~al.},\ }\href
  {https://www.nature.com/articles/s41566-018-0097-4} {\bibfield  {journal}
  {\bibinfo  {journal} {Nat. Photon.}\ }\textbf {\bibinfo {volume} {12}},\
  \bibinfo {pages} {173} (\bibinfo {year} {2018})}\BibitemShut {NoStop}%
\bibitem [{\citenamefont {Santagati}\ \emph {et~al.}(2018)\citenamefont
  {Santagati}, \citenamefont {Wang}, \citenamefont {Gentile}, \citenamefont
  {Paesani}, \citenamefont {Wiebe}, \citenamefont {McClean}, \citenamefont
  {Morley-Short}, \citenamefont {Shadbolt}, \citenamefont {Bonneau},
  \citenamefont {Silverstone}, \citenamefont {Tew}, \citenamefont {Zhou},
  \citenamefont {O{\textquoteright}Brien},\ and\ \citenamefont
  {Thompson}}]{Santagatieaap9646}%
  \BibitemOpen
  \bibfield  {author} {\bibinfo {author} {\bibfnamefont {R.}~\bibnamefont
  {Santagati}}, \bibinfo {author} {\bibfnamefont {J.}~\bibnamefont {Wang}},
  \bibinfo {author} {\bibfnamefont {A.~A.}\ \bibnamefont {Gentile}}, \bibinfo
  {author} {\bibfnamefont {S.}~\bibnamefont {Paesani}}, \bibinfo {author}
  {\bibfnamefont {N.}~\bibnamefont {Wiebe}}, \bibinfo {author} {\bibfnamefont
  {J.~R.}\ \bibnamefont {McClean}}, \bibinfo {author} {\bibfnamefont
  {S.}~\bibnamefont {Morley-Short}}, \bibinfo {author} {\bibfnamefont {P.~J.}\
  \bibnamefont {Shadbolt}}, \bibinfo {author} {\bibfnamefont {D.}~\bibnamefont
  {Bonneau}}, \bibinfo {author} {\bibfnamefont {J.~W.}\ \bibnamefont
  {Silverstone}}, \bibinfo {author} {\bibfnamefont {D.~P.}\ \bibnamefont
  {Tew}}, \bibinfo {author} {\bibfnamefont {X.}~\bibnamefont {Zhou}}, \bibinfo
  {author} {\bibfnamefont {J.~L.}\ \bibnamefont {O{\textquoteright}Brien}}, \
  and\ \bibinfo {author} {\bibfnamefont {M.~G.}\ \bibnamefont {Thompson}},\
  }\href {\doibase 10.1126/sciadv.aap9646} {\bibfield  {journal} {\bibinfo
  {journal} {Science Advances}\ }\textbf {\bibinfo {volume} {4}} (\bibinfo
  {year} {2018}),\ 10.1126/sciadv.aap9646},\ \Eprint
  {http://arxiv.org/abs/https://advances.sciencemag.org/content/4/1/eaap9646.full.pdf}
  {https://advances.sciencemag.org/content/4/1/eaap9646.full.pdf} \BibitemShut
  {NoStop}%
\bibitem [{\citenamefont {Agresti}\ \emph {et~al.}(2019)\citenamefont
  {Agresti}, \citenamefont {Viggianiello}, \citenamefont {Flamini},
  \citenamefont {Spagnolo}, \citenamefont {Crespi}, \citenamefont {Osellame},
  \citenamefont {Wiebe},\ and\ \citenamefont {Sciarrino}}]{agresti2019pattern}%
  \BibitemOpen
  \bibfield  {author} {\bibinfo {author} {\bibfnamefont {I.}~\bibnamefont
  {Agresti}}, \bibinfo {author} {\bibfnamefont {N.}~\bibnamefont
  {Viggianiello}}, \bibinfo {author} {\bibfnamefont {F.}~\bibnamefont
  {Flamini}}, \bibinfo {author} {\bibfnamefont {N.}~\bibnamefont {Spagnolo}},
  \bibinfo {author} {\bibfnamefont {A.}~\bibnamefont {Crespi}}, \bibinfo
  {author} {\bibfnamefont {R.}~\bibnamefont {Osellame}}, \bibinfo {author}
  {\bibfnamefont {N.}~\bibnamefont {Wiebe}}, \ and\ \bibinfo {author}
  {\bibfnamefont {F.}~\bibnamefont {Sciarrino}},\ }\href {\doibase
  10.1103/PhysRevX.9.011013} {\bibfield  {journal} {\bibinfo  {journal} {Phys.
  Rev. X}\ }\textbf {\bibinfo {volume} {9}},\ \bibinfo {pages} {011013}
  (\bibinfo {year} {2019})}\BibitemShut {NoStop}%
\bibitem [{\citenamefont {Lumino}\ \emph {et~al.}(2018)\citenamefont {Lumino},
  \citenamefont {Polino}, \citenamefont {Rab}, \citenamefont {Milani},
  \citenamefont {Spagnolo}, \citenamefont {Wiebe},\ and\ \citenamefont
  {Sciarrino}}]{lumino2018}%
  \BibitemOpen
  \bibfield  {author} {\bibinfo {author} {\bibfnamefont {A.}~\bibnamefont
  {Lumino}}, \bibinfo {author} {\bibfnamefont {E.}~\bibnamefont {Polino}},
  \bibinfo {author} {\bibfnamefont {A.~S.}\ \bibnamefont {Rab}}, \bibinfo
  {author} {\bibfnamefont {G.}~\bibnamefont {Milani}}, \bibinfo {author}
  {\bibfnamefont {N.}~\bibnamefont {Spagnolo}}, \bibinfo {author}
  {\bibfnamefont {N.}~\bibnamefont {Wiebe}}, \ and\ \bibinfo {author}
  {\bibfnamefont {F.}~\bibnamefont {Sciarrino}},\ }\href
  {https://journals.aps.org/prapplied/abstract/10.1103/PhysRevApplied.10.044033}
  {\bibfield  {journal} {\bibinfo  {journal} {Phys. Rev. Applied}\ }\textbf
  {\bibinfo {volume} {10}},\ \bibinfo {pages} {044033} (\bibinfo {year}
  {2018})}\BibitemShut {NoStop}%
\bibitem [{\citenamefont {Rocchetto}\ \emph {et~al.}(2019)\citenamefont
  {Rocchetto}, \citenamefont {Aaronson}, \citenamefont {Severini},
  \citenamefont {Carvacho}, \citenamefont {Poderini}, \citenamefont {Agresti},
  \citenamefont {Bentivegna},\ and\ \citenamefont {Sciarrino}}]{rocchetto2019}%
  \BibitemOpen
  \bibfield  {author} {\bibinfo {author} {\bibfnamefont {A.}~\bibnamefont
  {Rocchetto}}, \bibinfo {author} {\bibfnamefont {S.}~\bibnamefont {Aaronson}},
  \bibinfo {author} {\bibfnamefont {S.}~\bibnamefont {Severini}}, \bibinfo
  {author} {\bibfnamefont {G.}~\bibnamefont {Carvacho}}, \bibinfo {author}
  {\bibfnamefont {D.}~\bibnamefont {Poderini}}, \bibinfo {author}
  {\bibfnamefont {I.}~\bibnamefont {Agresti}}, \bibinfo {author} {\bibfnamefont
  {M.}~\bibnamefont {Bentivegna}}, \ and\ \bibinfo {author} {\bibfnamefont
  {F.}~\bibnamefont {Sciarrino}},\ }\href
  {https://advances.sciencemag.org/content/5/3/eaau1946} {\bibfield  {journal}
  {\bibinfo  {journal} {Science Advances}\ }\textbf {\bibinfo {volume} {5}},\
  \bibinfo {pages} {eaau1946} (\bibinfo {year} {2019})}\BibitemShut {NoStop}%
\bibitem [{\citenamefont {Butler}\ \emph {et~al.}(2018)\citenamefont {Butler},
  \citenamefont {Davies}, \citenamefont {Cartwright}, \citenamefont {Isayev},\
  and\ \citenamefont {Walsh}}]{butler2018}%
  \BibitemOpen
  \bibfield  {author} {\bibinfo {author} {\bibfnamefont {K.~T.}\ \bibnamefont
  {Butler}}, \bibinfo {author} {\bibfnamefont {D.~W.}\ \bibnamefont {Davies}},
  \bibinfo {author} {\bibfnamefont {H.}~\bibnamefont {Cartwright}}, \bibinfo
  {author} {\bibfnamefont {O.}~\bibnamefont {Isayev}}, \ and\ \bibinfo {author}
  {\bibfnamefont {A.}~\bibnamefont {Walsh}},\ }\href
  {https://www.nature.com/articles/s41586-018-0337-2} {\bibfield  {journal}
  {\bibinfo  {journal} {Nature}\ }\textbf {\bibinfo {volume} {559}},\ \bibinfo
  {pages} {547} (\bibinfo {year} {2018})}\BibitemShut {NoStop}%
\bibitem [{\citenamefont {Fischer}\ \emph {et~al.}(2006)\citenamefont
  {Fischer}, \citenamefont {Tibbetts}, \citenamefont {Morgan},\ and\
  \citenamefont {Ceder}}]{fischer2006}%
  \BibitemOpen
  \bibfield  {author} {\bibinfo {author} {\bibfnamefont {C.~C.}\ \bibnamefont
  {Fischer}}, \bibinfo {author} {\bibfnamefont {K.~J.}\ \bibnamefont
  {Tibbetts}}, \bibinfo {author} {\bibfnamefont {D.}~\bibnamefont {Morgan}}, \
  and\ \bibinfo {author} {\bibfnamefont {G.}~\bibnamefont {Ceder}},\ }\href
  {https://www.nature.com/articles/nmat1691} {\bibfield  {journal} {\bibinfo
  {journal} {Nat. Mater.}\ }\textbf {\bibinfo {volume} {5}},\ \bibinfo {pages}
  {641} (\bibinfo {year} {2006})}\BibitemShut {NoStop}%
\bibitem [{\citenamefont {Melnikov}\ \emph {et~al.}(2018)\citenamefont
  {Melnikov}, \citenamefont {Nautrup}, \citenamefont {Krenn}, \citenamefont
  {Dunjko}, \citenamefont {Tiersch}, \citenamefont {Zeilinger},\ and\
  \citenamefont {Briegel}}]{melnikov}%
  \BibitemOpen
  \bibfield  {author} {\bibinfo {author} {\bibfnamefont {A.~A.}\ \bibnamefont
  {Melnikov}}, \bibinfo {author} {\bibfnamefont {H.~P.}\ \bibnamefont
  {Nautrup}}, \bibinfo {author} {\bibfnamefont {M.}~\bibnamefont {Krenn}},
  \bibinfo {author} {\bibfnamefont {V.}~\bibnamefont {Dunjko}}, \bibinfo
  {author} {\bibfnamefont {M.}~\bibnamefont {Tiersch}}, \bibinfo {author}
  {\bibfnamefont {A.}~\bibnamefont {Zeilinger}}, \ and\ \bibinfo {author}
  {\bibfnamefont {H.~J.}\ \bibnamefont {Briegel}},\ }\href
  {https://www.pnas.org/content/115/6/1221.short} {\bibfield  {journal}
  {\bibinfo  {journal} {Proc. Nat. Acad. Sci. USA}\ }\textbf {\bibinfo {volume}
  {115}},\ \bibinfo {pages} {1221} (\bibinfo {year} {2018})}\BibitemShut
  {NoStop}%
\bibitem [{\citenamefont {Wang}\ \emph {et~al.}(2017)\citenamefont {Wang},
  \citenamefont {Paesani}, \citenamefont {Santagati}, \citenamefont {Knauer},
  \citenamefont {Gentile}, \citenamefont {Wiebe}, \citenamefont {Petruzzella},
  \citenamefont {O’Brien}, \citenamefont {Rarity}, \citenamefont {Laing}
  \emph {et~al.}}]{wangpaesani}%
  \BibitemOpen
  \bibfield  {author} {\bibinfo {author} {\bibfnamefont {J.}~\bibnamefont
  {Wang}}, \bibinfo {author} {\bibfnamefont {S.}~\bibnamefont {Paesani}},
  \bibinfo {author} {\bibfnamefont {R.}~\bibnamefont {Santagati}}, \bibinfo
  {author} {\bibfnamefont {S.}~\bibnamefont {Knauer}}, \bibinfo {author}
  {\bibfnamefont {A.~A.}\ \bibnamefont {Gentile}}, \bibinfo {author}
  {\bibfnamefont {N.}~\bibnamefont {Wiebe}}, \bibinfo {author} {\bibfnamefont
  {M.}~\bibnamefont {Petruzzella}}, \bibinfo {author} {\bibfnamefont {J.~L.}\
  \bibnamefont {O’Brien}}, \bibinfo {author} {\bibfnamefont {J.~G.}\
  \bibnamefont {Rarity}}, \bibinfo {author} {\bibfnamefont {A.}~\bibnamefont
  {Laing}},  \emph {et~al.},\ }\href
  {https://www.nature.com/articles/nphys4074?WT.feed_name=subjects_quantum-simulation}
  {\bibfield  {journal} {\bibinfo  {journal} {Nat. Phys.}\ }\textbf {\bibinfo
  {volume} {13}},\ \bibinfo {pages} {551} (\bibinfo {year} {2017})}\BibitemShut
  {NoStop}%
\bibitem [{\citenamefont {Cimini}\ \emph {et~al.}(2019)\citenamefont {Cimini},
  \citenamefont {Gianani}, \citenamefont {Spagnolo}, \citenamefont {Leccese},
  \citenamefont {Sciarrino},\ and\ \citenamefont {Barbieri}}]{Cimini2020}%
  \BibitemOpen
  \bibfield  {author} {\bibinfo {author} {\bibfnamefont {V.}~\bibnamefont
  {Cimini}}, \bibinfo {author} {\bibfnamefont {I.}~\bibnamefont {Gianani}},
  \bibinfo {author} {\bibfnamefont {N.}~\bibnamefont {Spagnolo}}, \bibinfo
  {author} {\bibfnamefont {F.}~\bibnamefont {Leccese}}, \bibinfo {author}
  {\bibfnamefont {F.}~\bibnamefont {Sciarrino}}, \ and\ \bibinfo {author}
  {\bibfnamefont {M.}~\bibnamefont {Barbieri}},\ }\href {\doibase
  10.1103/PhysRevLett.123.230502} {\bibfield  {journal} {\bibinfo  {journal}
  {Phys. Rev. Lett.}\ }\textbf {\bibinfo {volume} {123}},\ \bibinfo {pages}
  {230502} (\bibinfo {year} {2019})}\BibitemShut {NoStop}%
\bibitem [{\citenamefont {Krenn}\ \emph {et~al.}(2014)\citenamefont {Krenn},
  \citenamefont {Fickler}, \citenamefont {Fink}, \citenamefont {Handsteiner},
  \citenamefont {Malik}, \citenamefont {Scheidl}, \citenamefont {Ursin},\ and\
  \citenamefont {Zeilinger}}]{krenn_2014}%
  \BibitemOpen
  \bibfield  {author} {\bibinfo {author} {\bibfnamefont {M.}~\bibnamefont
  {Krenn}}, \bibinfo {author} {\bibfnamefont {R.}~\bibnamefont {Fickler}},
  \bibinfo {author} {\bibfnamefont {M.}~\bibnamefont {Fink}}, \bibinfo {author}
  {\bibfnamefont {J.}~\bibnamefont {Handsteiner}}, \bibinfo {author}
  {\bibfnamefont {M.}~\bibnamefont {Malik}}, \bibinfo {author} {\bibfnamefont
  {T.}~\bibnamefont {Scheidl}}, \bibinfo {author} {\bibfnamefont
  {R.}~\bibnamefont {Ursin}}, \ and\ \bibinfo {author} {\bibfnamefont
  {A.}~\bibnamefont {Zeilinger}},\ }\href {\doibase
  10.1088/1367-2630/16/11/113028} {\bibfield  {journal} {\bibinfo  {journal}
  {New J. Phys.}\ }\textbf {\bibinfo {volume} {16}},\ \bibinfo {pages} {113028}
  (\bibinfo {year} {2014})}\BibitemShut {NoStop}%
\bibitem [{\citenamefont {Krenn}\ \emph {et~al.}(2016)\citenamefont {Krenn},
  \citenamefont {Handsteiner}, \citenamefont {Fink}, \citenamefont {Fickler},
  \citenamefont {Ursin}, \citenamefont {Malik},\ and\ \citenamefont
  {Zeilinger}}]{krenn_2016}%
  \BibitemOpen
  \bibfield  {author} {\bibinfo {author} {\bibfnamefont {M.}~\bibnamefont
  {Krenn}}, \bibinfo {author} {\bibfnamefont {J.}~\bibnamefont {Handsteiner}},
  \bibinfo {author} {\bibfnamefont {M.}~\bibnamefont {Fink}}, \bibinfo {author}
  {\bibfnamefont {R.}~\bibnamefont {Fickler}}, \bibinfo {author} {\bibfnamefont
  {R.}~\bibnamefont {Ursin}}, \bibinfo {author} {\bibfnamefont
  {M.}~\bibnamefont {Malik}}, \ and\ \bibinfo {author} {\bibfnamefont
  {A.}~\bibnamefont {Zeilinger}},\ }\href {\doibase 10.1073/pnas.1612023113}
  {\bibfield  {journal} {\bibinfo  {journal} {Proc. Nat. Acad. Sci. USA}\
  }\textbf {\bibinfo {volume} {29}},\ \bibinfo {pages} {13648} (\bibinfo {year}
  {2016})}\BibitemShut {NoStop}%
\bibitem [{\citenamefont {Doster}\ and\ \citenamefont
  {Watnik}(2017)}]{Doster_17}%
  \BibitemOpen
  \bibfield  {author} {\bibinfo {author} {\bibfnamefont {T.}~\bibnamefont
  {Doster}}\ and\ \bibinfo {author} {\bibfnamefont {A.~T.}\ \bibnamefont
  {Watnik}},\ }\href {\doibase 10.1364/AO.56.003386} {\bibfield  {journal}
  {\bibinfo  {journal} {Appl. Opt.}\ }\textbf {\bibinfo {volume} {56}},\
  \bibinfo {pages} {3386} (\bibinfo {year} {2017})}\BibitemShut {NoStop}%
\bibitem [{\citenamefont {Park}\ \emph {et~al.}(2018)\citenamefont {Park},
  \citenamefont {Cattell}, \citenamefont {Nichols}, \citenamefont {Watnik},
  \citenamefont {Doster},\ and\ \citenamefont {Rohde}}]{Park_18}%
  \BibitemOpen
  \bibfield  {author} {\bibinfo {author} {\bibfnamefont {S.~R.}\ \bibnamefont
  {Park}}, \bibinfo {author} {\bibfnamefont {L.}~\bibnamefont {Cattell}},
  \bibinfo {author} {\bibfnamefont {J.~M.}\ \bibnamefont {Nichols}}, \bibinfo
  {author} {\bibfnamefont {A.}~\bibnamefont {Watnik}}, \bibinfo {author}
  {\bibfnamefont {T.}~\bibnamefont {Doster}}, \ and\ \bibinfo {author}
  {\bibfnamefont {G.~K.}\ \bibnamefont {Rohde}},\ }\href {\doibase
  10.1364/OE.26.004004} {\bibfield  {journal} {\bibinfo  {journal} {Opt.
  Express}\ }\textbf {\bibinfo {volume} {26}},\ \bibinfo {pages} {4004}
  (\bibinfo {year} {2018})}\BibitemShut {NoStop}%
\bibitem [{\citenamefont {Lohani}\ and\ \citenamefont
  {Glasser}(2018)}]{Lohani_turbo_18}%
  \BibitemOpen
  \bibfield  {author} {\bibinfo {author} {\bibfnamefont {S.}~\bibnamefont
  {Lohani}}\ and\ \bibinfo {author} {\bibfnamefont {R.~T.}\ \bibnamefont
  {Glasser}},\ }\href {\doibase 10.1364/OL.43.002611} {\bibfield  {journal}
  {\bibinfo  {journal} {Opt. Lett.}\ }\textbf {\bibinfo {volume} {43}},\
  \bibinfo {pages} {2611} (\bibinfo {year} {2018})}\BibitemShut {NoStop}%
\bibitem [{\citenamefont {Li}\ \emph {et~al.}(2018)\citenamefont {Li},
  \citenamefont {Zhang}, \citenamefont {Wang}, \citenamefont {Wu},\ and\
  \citenamefont {Zhan}}]{Li_18}%
  \BibitemOpen
  \bibfield  {author} {\bibinfo {author} {\bibfnamefont {J.}~\bibnamefont
  {Li}}, \bibinfo {author} {\bibfnamefont {M.}~\bibnamefont {Zhang}}, \bibinfo
  {author} {\bibfnamefont {D.}~\bibnamefont {Wang}}, \bibinfo {author}
  {\bibfnamefont {S.}~\bibnamefont {Wu}}, \ and\ \bibinfo {author}
  {\bibfnamefont {Y.}~\bibnamefont {Zhan}},\ }\href {\doibase
  10.1364/OE.26.010494} {\bibfield  {journal} {\bibinfo  {journal} {Opt.
  Express}\ }\textbf {\bibinfo {volume} {26}},\ \bibinfo {pages} {10494}
  (\bibinfo {year} {2018})}\BibitemShut {NoStop}%
\bibitem [{\citenamefont {Milione}\ \emph {et~al.}(2011)\citenamefont
  {Milione}, \citenamefont {Sztul}, \citenamefont {Nolan},\ and\ \citenamefont
  {Alfano}}]{milione_poinc_sphere_2011}%
  \BibitemOpen
  \bibfield  {author} {\bibinfo {author} {\bibfnamefont {G.}~\bibnamefont
  {Milione}}, \bibinfo {author} {\bibfnamefont {H.~I.}\ \bibnamefont {Sztul}},
  \bibinfo {author} {\bibfnamefont {D.~A.}\ \bibnamefont {Nolan}}, \ and\
  \bibinfo {author} {\bibfnamefont {R.~R.}\ \bibnamefont {Alfano}},\ }\href
  {\doibase 10.1103/PhysRevLett.107.053601} {\bibfield  {journal} {\bibinfo
  {journal} {Phys. Rev. Lett.}\ }\textbf {\bibinfo {volume} {107}},\ \bibinfo
  {pages} {053601} (\bibinfo {year} {2011})}\BibitemShut {NoStop}%
\bibitem [{\citenamefont {Cardano}\ \emph {et~al.}(2012)\citenamefont
  {Cardano}, \citenamefont {Karimi}, \citenamefont {Slussarenko}, \citenamefont
  {Marrucci}, \citenamefont {de~Lisio},\ and\ \citenamefont
  {Santamato}}]{Cardano:12}%
  \BibitemOpen
  \bibfield  {author} {\bibinfo {author} {\bibfnamefont {F.}~\bibnamefont
  {Cardano}}, \bibinfo {author} {\bibfnamefont {E.}~\bibnamefont {Karimi}},
  \bibinfo {author} {\bibfnamefont {S.}~\bibnamefont {Slussarenko}}, \bibinfo
  {author} {\bibfnamefont {L.}~\bibnamefont {Marrucci}}, \bibinfo {author}
  {\bibfnamefont {C.}~\bibnamefont {de~Lisio}}, \ and\ \bibinfo {author}
  {\bibfnamefont {E.}~\bibnamefont {Santamato}},\ }\href {\doibase
  10.1364/AO.51.0000C1} {\bibfield  {journal} {\bibinfo  {journal} {Appl.
  Opt.}\ }\textbf {\bibinfo {volume} {51}},\ \bibinfo {pages} {C1} (\bibinfo
  {year} {2012})}\BibitemShut {NoStop}%
\bibitem [{\citenamefont {Marrucci}\ \emph {et~al.}(2006)\citenamefont
  {Marrucci}, \citenamefont {Manzo},\ and\ \citenamefont
  {Paparo}}]{marrucci2006optical}%
  \BibitemOpen
  \bibfield  {author} {\bibinfo {author} {\bibfnamefont {L.}~\bibnamefont
  {Marrucci}}, \bibinfo {author} {\bibfnamefont {C.}~\bibnamefont {Manzo}}, \
  and\ \bibinfo {author} {\bibfnamefont {D.}~\bibnamefont {Paparo}},\ }\href
  {\doibase 10.1103/physrevlett.96.163905} {\bibfield  {journal} {\bibinfo
  {journal} {Phys. Rev. Lett.}\ }\textbf {\bibinfo {volume} {96}},\ \bibinfo
  {pages} {163905} (\bibinfo {year} {2006})}\BibitemShut {NoStop}%
\bibitem [{\citenamefont {LeCun}\ \emph {et~al.}(2015)\citenamefont {LeCun},
  \citenamefont {Bengio},\ and\ \citenamefont {Hinton}}]{lecun2015deep}%
  \BibitemOpen
  \bibfield  {author} {\bibinfo {author} {\bibfnamefont {Y.}~\bibnamefont
  {LeCun}}, \bibinfo {author} {\bibfnamefont {Y.}~\bibnamefont {Bengio}}, \
  and\ \bibinfo {author} {\bibfnamefont {G.}~\bibnamefont {Hinton}},\ }\href
  {\doibase 10.1038/nature14539} {\bibfield  {journal} {\bibinfo  {journal}
  {Nature}\ }\textbf {\bibinfo {volume} {521}},\ \bibinfo {pages} {436–444}
  (\bibinfo {year} {2015})}\BibitemShut {NoStop}%
\bibitem [{\citenamefont {Simard}\ \emph {et~al.}(2003)\citenamefont {Simard},
  \citenamefont {Steinkraus},\ and\ \citenamefont {Platt}}]{Simard2003}%
  \BibitemOpen
  \bibfield  {author} {\bibinfo {author} {\bibfnamefont {P.~Y.}\ \bibnamefont
  {Simard}}, \bibinfo {author} {\bibfnamefont {D.}~\bibnamefont {Steinkraus}},
  \ and\ \bibinfo {author} {\bibfnamefont {J.~C.}\ \bibnamefont {Platt}},\ }in\
  \href@noop {} {\emph {\bibinfo {booktitle} {ICDAR}}}\ (\bibinfo {year}
  {2003})\BibitemShut {NoStop}%
\bibitem [{\citenamefont {Cire\c{s}an}\ \emph {et~al.}(2011)\citenamefont
  {Cire\c{s}an}, \citenamefont {Meier}, \citenamefont {Masci}, \citenamefont
  {Gambardella},\ and\ \citenamefont
  {Schmidhuber}}]{Ciresan:2011:FHP:2283516.2283603}%
  \BibitemOpen
  \bibfield  {author} {\bibinfo {author} {\bibfnamefont {D.~C.}\ \bibnamefont
  {Cire\c{s}an}}, \bibinfo {author} {\bibfnamefont {U.}~\bibnamefont {Meier}},
  \bibinfo {author} {\bibfnamefont {J.}~\bibnamefont {Masci}}, \bibinfo
  {author} {\bibfnamefont {L.~M.}\ \bibnamefont {Gambardella}}, \ and\ \bibinfo
  {author} {\bibfnamefont {J.}~\bibnamefont {Schmidhuber}},\ }in\ \href
  {http://dx.doi.org/10.5591/978-1-57735-516-8/IJCAI11-210} {\emph {\bibinfo
  {booktitle} {Proceedings of the Twenty-Second International Joint Conference
  on Artificial Intelligence - Volume Volume Two}}},\ \bibinfo {series and
  number} {IJCAI'11}\ (\bibinfo  {publisher} {AAAI Press},\ \bibinfo {year}
  {2011})\ pp.\ \bibinfo {pages} {1237--1242}\BibitemShut {NoStop}%
\bibitem [{\citenamefont {Matsugu}\ \emph {et~al.}(2003)\citenamefont
  {Matsugu}, \citenamefont {Mori}, \citenamefont {Mitari},\ and\ \citenamefont
  {Kaneda}}]{MATSUGU2003555}%
  \BibitemOpen
  \bibfield  {author} {\bibinfo {author} {\bibfnamefont {M.}~\bibnamefont
  {Matsugu}}, \bibinfo {author} {\bibfnamefont {K.}~\bibnamefont {Mori}},
  \bibinfo {author} {\bibfnamefont {Y.}~\bibnamefont {Mitari}}, \ and\ \bibinfo
  {author} {\bibfnamefont {Y.}~\bibnamefont {Kaneda}},\ }\href
  {https://doi.org/10.1016/S0893-6080(03)00115-1} {\bibfield  {journal}
  {\bibinfo  {journal} {Neural Networks}\ }\textbf {\bibinfo {volume} {16}},\
  \bibinfo {pages} {555 } (\bibinfo {year} {2003})},\ \bibinfo {note} {advances
  in Neural Networks Research: IJCNN '03}\BibitemShut {NoStop}%
\bibitem [{SI()}]{SI}%
  \BibitemOpen
  \href@noop {} {}\bibinfo {note} {The Supplemental Material accompanying the
  paper, available from XXXX, provides further details of the experimental
  setup an the machine learning techniques, and includes Refs.
  \cite{cardano2015quantum,Innocenti2017,giordani_2018,milione_poinc_sphere_2011,Cardano:12,marrucci2006optical,github,chollet2015keras,tensorflow2015-whitepaper,ruder2016overview,karimi_07,karimi_09,cunningham2008dimension,fodor2002survey,jolliffe2016principal,Hearst_si,cristianini_shawe-taylor_2000}}\BibitemShut
  {NoStop}%
\bibitem [{git()}]{github}%
  \BibitemOpen
  \href@noop {} {}\bibinfo {howpublished} {The source code and the dataset for
  reproducing the results of this work are available at
  \url{https://github.com/lucainnocenti/ML-classification-of-VVBs}}\BibitemShut
  {NoStop}%
\bibitem [{\citenamefont {Chollet}\ \emph {et~al.}(2015)\citenamefont {Chollet}
  \emph {et~al.}}]{chollet2015keras}%
  \BibitemOpen
  \bibfield  {author} {\bibinfo {author} {\bibfnamefont {F.}~\bibnamefont
  {Chollet}} \emph {et~al.},\ }\href@noop {} {\enquote {\bibinfo {title}
  {Keras},}\ }\bibinfo {howpublished} {\url{https://keras.io}} (\bibinfo {year}
  {2015})\BibitemShut {NoStop}%
\bibitem [{\citenamefont {Abadi}\ \emph {et~al.}(2015)\citenamefont {Abadi},
  \citenamefont {Agarwal}, \citenamefont {Barham}, \citenamefont {Brevdo},
  \citenamefont {Chen}, \citenamefont {Citro}, \citenamefont {Corrado},
  \citenamefont {Davis}, \citenamefont {Dean}, \citenamefont {Devin},
  \citenamefont {Ghemawat}, \citenamefont {Goodfellow}, \citenamefont {Harp},
  \citenamefont {Irving}, \citenamefont {Isard}, \citenamefont {Jia},
  \citenamefont {Jozefowicz}, \citenamefont {Kaiser}, \citenamefont {Kudlur},
  \citenamefont {Levenberg}, \citenamefont {Man\'{e}}, \citenamefont {Monga},
  \citenamefont {Moore}, \citenamefont {Murray}, \citenamefont {Olah},
  \citenamefont {Schuster}, \citenamefont {Shlens}, \citenamefont {Steiner},
  \citenamefont {Sutskever}, \citenamefont {Talwar}, \citenamefont {Tucker},
  \citenamefont {Vanhoucke}, \citenamefont {Vasudevan}, \citenamefont
  {Vi\'{e}gas}, \citenamefont {Vinyals}, \citenamefont {Warden}, \citenamefont
  {Wattenberg}, \citenamefont {Wicke}, \citenamefont {Yu},\ and\ \citenamefont
  {Zheng}}]{tensorflow2015-whitepaper}%
  \BibitemOpen
  \bibfield  {author} {\bibinfo {author} {\bibfnamefont {M.}~\bibnamefont
  {Abadi}}, \bibinfo {author} {\bibfnamefont {A.}~\bibnamefont {Agarwal}},
  \bibinfo {author} {\bibfnamefont {P.}~\bibnamefont {Barham}}, \bibinfo
  {author} {\bibfnamefont {E.}~\bibnamefont {Brevdo}}, \bibinfo {author}
  {\bibfnamefont {Z.}~\bibnamefont {Chen}}, \bibinfo {author} {\bibfnamefont
  {C.}~\bibnamefont {Citro}}, \bibinfo {author} {\bibfnamefont {G.~S.}\
  \bibnamefont {Corrado}}, \bibinfo {author} {\bibfnamefont {A.}~\bibnamefont
  {Davis}}, \bibinfo {author} {\bibfnamefont {J.}~\bibnamefont {Dean}},
  \bibinfo {author} {\bibfnamefont {M.}~\bibnamefont {Devin}}, \bibinfo
  {author} {\bibfnamefont {S.}~\bibnamefont {Ghemawat}}, \bibinfo {author}
  {\bibfnamefont {I.}~\bibnamefont {Goodfellow}}, \bibinfo {author}
  {\bibfnamefont {A.}~\bibnamefont {Harp}}, \bibinfo {author} {\bibfnamefont
  {G.}~\bibnamefont {Irving}}, \bibinfo {author} {\bibfnamefont
  {M.}~\bibnamefont {Isard}}, \bibinfo {author} {\bibfnamefont
  {Y.}~\bibnamefont {Jia}}, \bibinfo {author} {\bibfnamefont {R.}~\bibnamefont
  {Jozefowicz}}, \bibinfo {author} {\bibfnamefont {L.}~\bibnamefont {Kaiser}},
  \bibinfo {author} {\bibfnamefont {M.}~\bibnamefont {Kudlur}}, \bibinfo
  {author} {\bibfnamefont {J.}~\bibnamefont {Levenberg}}, \bibinfo {author}
  {\bibfnamefont {D.}~\bibnamefont {Man\'{e}}}, \bibinfo {author}
  {\bibfnamefont {R.}~\bibnamefont {Monga}}, \bibinfo {author} {\bibfnamefont
  {S.}~\bibnamefont {Moore}}, \bibinfo {author} {\bibfnamefont
  {D.}~\bibnamefont {Murray}}, \bibinfo {author} {\bibfnamefont
  {C.}~\bibnamefont {Olah}}, \bibinfo {author} {\bibfnamefont {M.}~\bibnamefont
  {Schuster}}, \bibinfo {author} {\bibfnamefont {J.}~\bibnamefont {Shlens}},
  \bibinfo {author} {\bibfnamefont {B.}~\bibnamefont {Steiner}}, \bibinfo
  {author} {\bibfnamefont {I.}~\bibnamefont {Sutskever}}, \bibinfo {author}
  {\bibfnamefont {K.}~\bibnamefont {Talwar}}, \bibinfo {author} {\bibfnamefont
  {P.}~\bibnamefont {Tucker}}, \bibinfo {author} {\bibfnamefont
  {V.}~\bibnamefont {Vanhoucke}}, \bibinfo {author} {\bibfnamefont
  {V.}~\bibnamefont {Vasudevan}}, \bibinfo {author} {\bibfnamefont
  {F.}~\bibnamefont {Vi\'{e}gas}}, \bibinfo {author} {\bibfnamefont
  {O.}~\bibnamefont {Vinyals}}, \bibinfo {author} {\bibfnamefont
  {P.}~\bibnamefont {Warden}}, \bibinfo {author} {\bibfnamefont
  {M.}~\bibnamefont {Wattenberg}}, \bibinfo {author} {\bibfnamefont
  {M.}~\bibnamefont {Wicke}}, \bibinfo {author} {\bibfnamefont
  {Y.}~\bibnamefont {Yu}}, \ and\ \bibinfo {author} {\bibfnamefont
  {X.}~\bibnamefont {Zheng}},\ }\href {https://www.tensorflow.org/} {\enquote
  {\bibinfo {title} {{TensorFlow}: Large-scale machine learning on
  heterogeneous systems},}\ } (\bibinfo {year} {2015}),\ \bibinfo {note}
  {software available from tensorflow.org}\BibitemShut {NoStop}%
\bibitem [{\citenamefont {Ruder}(2016)}]{ruder2016overview}%
  \BibitemOpen
  \bibfield  {author} {\bibinfo {author} {\bibfnamefont {S.}~\bibnamefont
  {Ruder}},\ }\href {http://arxiv.org/abs/1609.04747v2} {\  (\bibinfo {year}
  {2016})},\ \Eprint {http://arxiv.org/abs/1609.04747} {arXiv:1609.04747
  [cs.LG]} \BibitemShut {NoStop}%
\bibitem [{\citenamefont {Karimi}\ \emph {et~al.}(2007)\citenamefont {Karimi},
  \citenamefont {Zito}, \citenamefont {Piccirillo}, \citenamefont {Marrucci},\
  and\ \citenamefont {Santamato}}]{karimi_07}%
  \BibitemOpen
  \bibfield  {author} {\bibinfo {author} {\bibfnamefont {E.}~\bibnamefont
  {Karimi}}, \bibinfo {author} {\bibfnamefont {G.}~\bibnamefont {Zito}},
  \bibinfo {author} {\bibfnamefont {B.}~\bibnamefont {Piccirillo}}, \bibinfo
  {author} {\bibfnamefont {L.}~\bibnamefont {Marrucci}}, \ and\ \bibinfo
  {author} {\bibfnamefont {E.}~\bibnamefont {Santamato}},\ }\href {\doibase
  10.1364/OL.32.003053} {\bibfield  {journal} {\bibinfo  {journal} {Opt.
  Lett.}\ }\textbf {\bibinfo {volume} {32}},\ \bibinfo {pages} {3053} (\bibinfo
  {year} {2007})}\BibitemShut {NoStop}%
\bibitem [{\citenamefont {Karimi}\ \emph {et~al.}(2009)\citenamefont {Karimi},
  \citenamefont {Piccirillo}, \citenamefont {Marrucci},\ and\ \citenamefont
  {Santamato}}]{karimi_09}%
  \BibitemOpen
  \bibfield  {author} {\bibinfo {author} {\bibfnamefont {E.}~\bibnamefont
  {Karimi}}, \bibinfo {author} {\bibfnamefont {B.}~\bibnamefont {Piccirillo}},
  \bibinfo {author} {\bibfnamefont {L.}~\bibnamefont {Marrucci}}, \ and\
  \bibinfo {author} {\bibfnamefont {E.}~\bibnamefont {Santamato}},\ }\href
  {\doibase 10.1364/OL.34.001225} {\bibfield  {journal} {\bibinfo  {journal}
  {Opt. Lett.}\ }\textbf {\bibinfo {volume} {34}},\ \bibinfo {pages} {1225}
  (\bibinfo {year} {2009})}\BibitemShut {NoStop}%
\bibitem [{\citenamefont {Rafayelyan}\ \emph {et~al.}(2017)\citenamefont
  {Rafayelyan}, \citenamefont {Gertus},\ and\ \citenamefont
  {Brasselet}}]{Rafayelyan}%
  \BibitemOpen
  \bibfield  {author} {\bibinfo {author} {\bibfnamefont {M.}~\bibnamefont
  {Rafayelyan}}, \bibinfo {author} {\bibfnamefont {T.}~\bibnamefont {Gertus}},
  \ and\ \bibinfo {author} {\bibfnamefont {E.}~\bibnamefont {Brasselet}},\
  }\href {\doibase 10.1063/1.4990954} {\bibfield  {journal} {\bibinfo
  {journal} {Applied Physics Letters}\ }\textbf {\bibinfo {volume} {110}},\
  \bibinfo {pages} {261108} (\bibinfo {year} {2017})},\ \Eprint
  {http://arxiv.org/abs/https://doi.org/10.1063/1.499095}
  {https://doi.org/10.1063/1.499095} \BibitemShut {NoStop}%
\bibitem [{\citenamefont {Shu}\ \emph {et~al.}(2016)\citenamefont {Shu},
  \citenamefont {Liu}, \citenamefont {Ke}, \citenamefont {Ling}, \citenamefont
  {Liu}, \citenamefont {Huang}, \citenamefont {Luo},\ and\ \citenamefont
  {Yin}}]{Shu:16}%
  \BibitemOpen
  \bibfield  {author} {\bibinfo {author} {\bibfnamefont {W.}~\bibnamefont
  {Shu}}, \bibinfo {author} {\bibfnamefont {Y.}~\bibnamefont {Liu}}, \bibinfo
  {author} {\bibfnamefont {Y.}~\bibnamefont {Ke}}, \bibinfo {author}
  {\bibfnamefont {X.}~\bibnamefont {Ling}}, \bibinfo {author} {\bibfnamefont
  {Z.}~\bibnamefont {Liu}}, \bibinfo {author} {\bibfnamefont {B.}~\bibnamefont
  {Huang}}, \bibinfo {author} {\bibfnamefont {H.}~\bibnamefont {Luo}}, \ and\
  \bibinfo {author} {\bibfnamefont {X.}~\bibnamefont {Yin}},\ }\href {\doibase
  10.1364/OE.24.021177} {\bibfield  {journal} {\bibinfo  {journal} {Opt.
  Express}\ }\textbf {\bibinfo {volume} {24}},\ \bibinfo {pages} {21177}
  (\bibinfo {year} {2016})}\BibitemShut {NoStop}%
\bibitem [{\citenamefont {Vallone}\ \emph {et~al.}(2016)\citenamefont
  {Vallone}, \citenamefont {Sponselli}, \citenamefont {D'Ambrosio},
  \citenamefont {Marrucci}, \citenamefont {Sciarrino},\ and\ \citenamefont
  {Villoresi}}]{Vallone:16}%
  \BibitemOpen
  \bibfield  {author} {\bibinfo {author} {\bibfnamefont {G.}~\bibnamefont
  {Vallone}}, \bibinfo {author} {\bibfnamefont {A.}~\bibnamefont {Sponselli}},
  \bibinfo {author} {\bibfnamefont {V.}~\bibnamefont {D'Ambrosio}}, \bibinfo
  {author} {\bibfnamefont {L.}~\bibnamefont {Marrucci}}, \bibinfo {author}
  {\bibfnamefont {F.}~\bibnamefont {Sciarrino}}, \ and\ \bibinfo {author}
  {\bibfnamefont {P.}~\bibnamefont {Villoresi}},\ }\href {\doibase
  10.1364/OE.24.016390} {\bibfield  {journal} {\bibinfo  {journal} {Opt.
  Express}\ }\textbf {\bibinfo {volume} {24}},\ \bibinfo {pages} {16390}
  (\bibinfo {year} {2016})}\BibitemShut {NoStop}%
\bibitem [{\citenamefont {Cunningham}(2008)}]{cunningham2008dimension}%
  \BibitemOpen
  \bibfield  {author} {\bibinfo {author} {\bibfnamefont {P.}~\bibnamefont
  {Cunningham}},\ }in\ \href {https://doi.org/10.1007/978-3-540-75171-7} {\emph
  {\bibinfo {booktitle} {Machine learning techniques for multimedia}}}\
  (\bibinfo  {publisher} {Springer},\ \bibinfo {year} {2008})\ pp.\ \bibinfo
  {pages} {91--112}\BibitemShut {NoStop}%
\bibitem [{\citenamefont {Fodor}(2002)}]{fodor2002survey}%
  \BibitemOpen
  \bibfield  {author} {\bibinfo {author} {\bibfnamefont {I.~K.}\ \bibnamefont
  {Fodor}},\ }\href {\doibase 10.2172/15002155} {\emph {\bibinfo {title} {A
  survey of dimension reduction techniques}}},\ \bibinfo {type} {Tech. Rep.}\
  (\bibinfo  {institution} {Lawrence Livermore National Lab., CA (US)},\
  \bibinfo {year} {2002})\BibitemShut {NoStop}%
\bibitem [{\citenamefont {Jolliffe}(2011)}]{jolliffe2011principal}%
  \BibitemOpen
  \bibfield  {author} {\bibinfo {author} {\bibfnamefont {I.}~\bibnamefont
  {Jolliffe}},\ }\href@noop {} {\emph {\bibinfo {title} {Principal component
  analysis}}}\ (\bibinfo  {publisher} {Springer},\ \bibinfo {year}
  {2011})\BibitemShut {NoStop}%
\bibitem [{\citenamefont {Jolliffe}\ and\ \citenamefont
  {Cadima}(2016)}]{jolliffe2016principal}%
  \BibitemOpen
  \bibfield  {author} {\bibinfo {author} {\bibfnamefont {I.~T.}\ \bibnamefont
  {Jolliffe}}\ and\ \bibinfo {author} {\bibfnamefont {J.}~\bibnamefont
  {Cadima}},\ }\href@noop {} {\bibfield  {journal} {\bibinfo  {journal}
  {Philosophical Transactions of the Royal Society A: Mathematical, Physical
  and Engineering Sciences}\ }\textbf {\bibinfo {volume} {374}},\ \bibinfo
  {pages} {20150202} (\bibinfo {year} {2016})}\BibitemShut {NoStop}%
\bibitem [{\citenamefont {{Hearst}}\ \emph {et~al.}(1998)\citenamefont
  {{Hearst}}, \citenamefont {{Dumais}}, \citenamefont {{Osuna}}, \citenamefont
  {{Platt}},\ and\ \citenamefont {{Scholkopf}}}]{Hearst_si}%
  \BibitemOpen
  \bibfield  {author} {\bibinfo {author} {\bibfnamefont {M.~A.}\ \bibnamefont
  {{Hearst}}}, \bibinfo {author} {\bibfnamefont {S.~T.}\ \bibnamefont
  {{Dumais}}}, \bibinfo {author} {\bibfnamefont {E.}~\bibnamefont {{Osuna}}},
  \bibinfo {author} {\bibfnamefont {J.}~\bibnamefont {{Platt}}}, \ and\
  \bibinfo {author} {\bibfnamefont {B.}~\bibnamefont {{Scholkopf}}},\ }\href
  {\doibase 10.1109/5254.708428} {\bibfield  {journal} {\bibinfo  {journal}
  {IEEE Intelligent Systems and their Applications}\ }\textbf {\bibinfo
  {volume} {13}},\ \bibinfo {pages} {18} (\bibinfo {year} {1998})}\BibitemShut
  {NoStop}%
\bibitem [{\citenamefont {Cristianini}\ and\ \citenamefont
  {Shawe-Taylor}(2000)}]{cristianini_shawe-taylor_2000}%
  \BibitemOpen
  \bibfield  {author} {\bibinfo {author} {\bibfnamefont {N.}~\bibnamefont
  {Cristianini}}\ and\ \bibinfo {author} {\bibfnamefont {J.}~\bibnamefont
  {Shawe-Taylor}},\ }\href {\doibase 10.1017/CBO9780511801389} {\emph {\bibinfo
  {title} {An Introduction to Support Vector Machines and Other Kernel-based
  Learning Methods}}}\ (\bibinfo  {publisher} {Cambridge University Press},\
  \bibinfo {year} {2000})\BibitemShut {NoStop}%
\bibitem [{\citenamefont {Liu}\ \emph {et~al.}(2019)\citenamefont {Liu},
  \citenamefont {Yan}, \citenamefont {Liu},\ and\ \citenamefont
  {Chen}}]{Zhanwei_oamCNN}%
  \BibitemOpen
  \bibfield  {author} {\bibinfo {author} {\bibfnamefont {Z.}~\bibnamefont
  {Liu}}, \bibinfo {author} {\bibfnamefont {S.}~\bibnamefont {Yan}}, \bibinfo
  {author} {\bibfnamefont {H.}~\bibnamefont {Liu}}, \ and\ \bibinfo {author}
  {\bibfnamefont {X.}~\bibnamefont {Chen}},\ }\href {\doibase
  10.1103/PhysRevLett.123.183902} {\bibfield  {journal} {\bibinfo  {journal}
  {Phys. Rev. Lett.}\ }\textbf {\bibinfo {volume} {123}},\ \bibinfo {pages}
  {183902} (\bibinfo {year} {2019})}\BibitemShut {NoStop}%
\end{thebibliography}
\end{document}